\documentclass[%
aip,
jcp,
amsmath,amssymb,
preprint,%
]{revtex4-2}

\usepackage{graphicx}
\usepackage{dcolumn}
\usepackage{bm}

\usepackage[utf8]{inputenc}
\usepackage[T1]{fontenc}
\usepackage{mathptmx}
\usepackage{etoolbox}

\usepackage{comment}

\usepackage{enumerate}
\usepackage{algorithm}
\usepackage{algorithmicx}
\usepackage{algpseudocode}
\usepackage{paralist}
\usepackage{threeparttable}
\usepackage{threeparttablex}
\usepackage{booktabs}
\usepackage{makecell}
\usepackage{multirow}
\usepackage{color}
\usepackage[mathscr]{eucal}
\usepackage{amsthm}
\usepackage{gensymb}
\usepackage{pifont}

\newcommand{\dcite}[1]{\scalebox{1.3}[1.3]{\raisebox{-0.65ex}{\cite{#1}}}}
\newtheorem{prop}{Proposition}

\algnewcommand{\Inputs}[1]{
  \State \textbf{Inputs:}
  \Statex \hspace*{\algorithmicindent}\parbox[t]{.8\linewidth}{\raggedright #1}
}
\algnewcommand{\Initialize}[1]{
  \State \textbf{Initialize:}
  \Statex \hspace*{\algorithmicindent}\parbox[t]{.8\linewidth}{\raggedright #1}
}
\algnewcommand{\Outputs}[1]{
  \State \textbf{Outputs:}
  \Statex \hspace*{\algorithmicindent}\parbox[t]{.8\linewidth}{\raggedright #1}
}
\algnewcommand\algorithmicswitch{\textbf{switch}}
\algnewcommand\algorithmiccase{\textbf{case}}
\algdef{SE}[SWITCH]{Switch}{EndSwitch}[1]{\algorithmicswitch\ #1\ \algorithmicdo}{\algorithmicend\ \algorithmicswitch}%
\algdef{SE}[CASE]{Case}{EndCase}[1]{\algorithmiccase\ #1}{\algorithmicend\ \algorithmiccase}%
\algtext*{EndSwitch}%
\algtext*{EndCase}%

\newfloat{Algorithm}{htbp}{alg}
\floatname{Algorithm}{Algorithm}

\makeatletter
\def\@email#1#2{%
 \endgroup
 \patchcmd{\titleblock@produce}
  {\frontmatter@RRAPformat}
  {\frontmatter@RRAPformat{\produce@RRAP{*#1\href{mailto:#2}{#2}}}\frontmatter@RRAPformat}
  {}{}
}%
\makeatother

\UseRawInputEncoding
\begin{document}

\preprint{}

\title[Kinetic network in Milestoning: Clustering, reduction, and transition path analysis]{Kinetic network in Milestoning: Clustering, reduction, and transition path analysis}

\author{Ru Wang}
\affiliation{
Qingdao Institute for Theoretical and Computational Sciences, School of Chemistry and Chemical Engineering, Shandong University, Qingdao, Shandong 266237, P. R. China}

\author{Xiaojun Ji*}
\email{jixj@sdu.edu.cn}
\affiliation{Research Center for Mathematics and Interdisciplinary Sciences, Shandong University, Qingdao, Shandong 266237, P. R. China}
\affiliation{Frontiers Science Center for Nonlinear Expectations (Ministry of Education), Shandong University, Qingdao, Shandong 266237, P. R. China}

\author{Hao Wang*}
\email{wanghaosd@sdu.edu.cn}
\affiliation{
Qingdao Institute for Theoretical and Computational Sciences, School of Chemistry and Chemical Engineering, Shandong University, Qingdao, Shandong 266237, P. R. China}

\author{Wenjian Liu}
\affiliation{
Qingdao Institute for Theoretical and Computational Sciences, School of Chemistry and Chemical Engineering, Shandong University, Qingdao, Shandong 266237, P. R. China}


\begin{abstract}
We present a reduction of Milestoning (ReM) algorithm to analyze the high-dimensional Milestoning kinetic network.
The algorithm reduces the Milestoning network to low dimensions but preserves essential kinetic information, such as local residence time, exit time, and mean first passage time between any two states. 
This is achieved in three steps.
First, nodes (milestones) in the high-dimensional Milestoning network are grouped into clusters based on the metastability identified by an auxiliary continuous-time Markov chain. 
Our clustering method is applicable not only to time-reversible networks but also to non-reversible networks generated from practical simulations with statistical fluctuations.
Second, a reduced network is established via network transformation, containing only the core sets of clusters as nodes.
Finally, transition pathways are analyzed in the reduced network based on the transition path theory.
The algorithm is illustrated using a toy model and a solvated alanine dipeptide in two and four dihedral angles.
\end{abstract}

\maketitle

\onecolumngrid

\section{Introduction}
The use of stochastic transition models in investigations of rare transition processes between metastable states in complex systems, such as protein folding and protein-ligand unbinding, has gained great popularity in recent years.
Continuous motion in the phase space is mapped into stochastic transitions among discretized states, as employed in methods such as Markov State Model\cite{MSM11,MSM14,MSM18} (MSM), Weighted Ensemble\cite{WE96,WE10} (WE), and Milestoning\cite{CM04,RevMile21}.
These discretized states are typically defined in the feature space or collective variable space and provide a coarse-grained description of the rare transition processes.

The stochastic transition model built on discretized states naturally induces a kinetic network, where the discretized states serve as nodes and the transition probabilities or rates among them represent edges.
This network contains rich information on the underlying mechanism, and simplified reaction coordinates can be efficiently extracted to provide intuitive interpretations. 
For example, insightful quantities such as the committor function\cite{CommMile,ComminMile} and the exit time to the product\cite{TranTime} ($\tau_P^e(x)$) can be conveniently estimated in the Milestoning formulation.
The committor function is considered as the optimal one-dimensional reaction coordinate. 
It is the probability that a trajectory starting at a phase space point will first reach the product state before the reactant state.
But one potential drawback of using the committor function as the reaction coordinate is the lack of time information.
The exit time to product provides an alternative definition of the reaction coordinate that includes time information explicitly. 
It measures the time that a trajectory starting at a phase space point takes to first reach the product state before the reactant state. 
Recently, a transition function ($\ln(\tau_P^e(x)/\tau_R^e(x))$) has also been proposed as a one-dimensional reaction coordinate containing temporal properties\cite{TranFunc}, which treats the time to the reactant and product states on an equal footing.

Compared to these simplified one-dimensional reaction coordinates, reaction pathways provide a more detailed description of the transition process.
In particular, the dominant reaction pathway offers a representative picture of the typical transition process.
Reaction pathways can be efficiently analyzed in kinetic networks.
For example, a recursive algorithm has been proposed for finding the dominant reaction pathway in a continuous-time Markov chain (CTMC) using the transition path theory (TPT)\cite{TPTMJ}. 
Reaction pathways in various network representations of kinetic data in the Milestoning formulation have also been carefully investigated in Ref. \dcite{MileNet}.
However, as the network generated from molecular dynamics (MD) simulations becomes increasingly complex, intuitive interpretation and qualitative understanding of transition pathways are made more difficult.
To address this issue, it is desirable to simplify the network structure while preserving the kinetic properties.

Dimensional reduction of a kinetic network can be performed in different ways\cite{OptRMSM,RGMSM,GT06,GT09}. 
Here, nodes in the original high-dimensional network are first clustered based on the metastability (timescale separation) of the system, and a reduced network consisting of only the core sets of clusters is then established.
For a Markov network described by a time-reversible transition rate matrix $\mathbf{Q}$, metastability is indicated by the spectral gap in the real-valued eigenvalues of $\mathbf{Q}$.
The right eigenvectors (the sum of each row in $\mathbf{Q}$ is zero) corresponding to those eigenvalues close to zero contain characteristic information of metastable states.
However, due to statistical fluctuations, this property may not hold even if the underlying dynamics is time-reversible.
As a result, the eigenvalues and eigenvectors of $\mathbf{Q}$ may become complex-valued, which makes the ranking of eigenvalues intractable.
To circumvent this issue, one possible solution is to correct $\mathbf{Q}$ (or equivalently, the transition probability matrix in the discrete-time case like MSM) by enforcing the detailed balance condition\cite{EnfDB04,EnfDB06,MSM09,RMSM}.

The situation is more complex for a non-Markovian network such as that of Milestoning\cite{MathExM16}.
Fortunately, an auxiliary CTMC that shares the same stationary probability, local residence time, and mean first passage time (MFPT) between any two nodes can be constructed\cite{Mile19}, providing a basis for clustering.
In practical scenarios, this auxiliary CTMC may also be non-reversible.
To address this issue, we propose an alternative method by introducing a composite matrix $\mathbf{A}=\mathbf{Q}\tilde{\mathbf{Q}}$, where $\tilde{\mathbf{Q}}$ is the time-reversed transition rate matrix.
This composite matrix $\mathbf{A}$ has several nice properties (see below for details), including satisfying the detailed balance condition, and is expected to preserve the metastability when the deviation from time-reversibility due to statistical fluctuations is small.
Consequently, the clustering is performed based on the eigenvalues and right eigenvectors of the composite matrix $\mathbf{A}$.
After clustering, the core set of each cluster is determined based on the magnitudes of free energies.
Subsequently, in the reduced network consisting of only the core sets as nodes, transition probabilities and transition time are recalculated such that the kinetics is preserved. 
Finally, transition pathways are examined in the reduced network.
This reduction analysis algorithm is dubbed ReM. See Algorithm \ref{ReM} for the pseudocode.

The remainder of this paper is organized as follows. 
In Sec. \ref{Methods}, we first briefly review the Milestoning formulation and then introduce the details of the ReM algorithm. 
Next, in Sec. \ref{Simulation}, computational details of a two-dimensional model system and a solvated alanine dipeptide are summarized. 
In Sec. \ref{Results}, the ReM algorithm is illustrated on the two-dimensional model system and the solvated alanine dipeptide in two and four dihedral angles. 
Finally, Sec. \ref{Conclusion} contains the concluding remarks.

\section{Method}\label{Methods}
\subsection{Milestoning Method}\label{Milestoning}
The basic idea of Milestoning is to integrate local fluxes to predict global kinetics\cite{CM04,ExM15}.
To this end, the phase space is divided into small cells, e.g., by Voronoi tessellation.
Interfaces between cells are called milestones, denoted by $\mathcal{M}=\{a,b,c,\cdots\}$.
The total number of milestones is assumed to be $N$. 

The governing equation in Milestoning is given by\cite{ExM15}
\begin{equation}
q_a(t)=p_a(0)\delta(t)+\sum_{b\in\mathcal{M}}\int_0^tq_b(t')K_{ba}(t-t') dt',
\label{basic eq}
\end{equation}
where $q_a(t)$ is the flux through milestone $a$ at time $t$, $p_a(0)$ is the initial probability of being at milestone $a$, $\delta(t)$ is the Dirac delta function, and $K_{ba}(t)$ is the transition kernel from milestone $b$ to $a$ ($a$ and $b$ are adjacent) with a time duration of $t$. 
Equation \eqref{basic eq} simply states the fact that the flux is conserved throughout the entire phase space, as there is no "sink" or "source".

The normalization condition of $K_{ba}(t)$ is as follows
\begin{equation}
\int_0^\infty K_{ba}(t) dt=K_{ba},
\label{K def}
\end{equation}
where $K_{ba}$ denotes the transition probability from milestone $b$ to $a$, and $\sum_{a\in\mathcal{M}}K_{ba}=1$.
In other words, the transition kernel can be written as
\begin{equation}
K_{ba}(t)=K_{ba}f_{ba}(t),
\end{equation}
where $f_{ba}(t)$ is the normalized transition time distribution from milestone $b$ to $a$.

Another useful quantity in Milestoning is the unnormalized transition time from milestone $b$ to $a$, which is defined as
\begin{equation}
T_{ba}=\int_0^\infty tK_{ba}(t)dt.
\label{T def}
\end{equation} 
A closely related quantity is the average residence time on milestone $b$,
\begin{equation}
t_b = \sum_{a\in\mathcal{M}} T_{ba},
\end{equation}
or in matrix notation,
\begin{equation}
\mathbf{t}=\mathbf{T}\mathbf{1}_N,
\end{equation}
with $\mathbf{1}_N$ is a column vector of length $N$, $\mathbf{1}^T_N=(1,\cdots,1)$.

In practical Milestoning calculations, the transition probability matrix $\mathbf{K}$ and unnormalized transition time matrix $\mathbf{T}$ are calculated by short trajectory simulations.
The initial distribution of these short trajectories determines the systematic error of $\mathbf{K}$ and $\mathbf{T}$.
The true initial distribution should be the first hitting point distribution (FHPD), which has no analytic expression in the general case\cite{ExM15}.
The local passage time weighted Milestoning (LPT-M) method has been demonstrated to provide an accurate and efficient estimation of FHPD\cite{LPTM} and is therefore adopted in this paper.
In LPT-M, initial configurations are sampled on each milestone using restraints according to the Boltzmann distribution.
Subsequently, free trajectories are evolved from these configurations both forward and backward in time until they hit a different milestone.
Finally, the trajectory weight is set inversely proportional to its local passage time through a infinitesimal interval around the initially sampled milestone.

Consequently, the transition probability and the unnormalized transition time are estimated as
\begin{equation}
K_{ba}=\frac{\sum_{l=1}^{n_b}\delta_{i(l)a}w(l)}{\sum_{l=1}^{n_b}w(l)},
\label{K simu}
\end{equation} 
\begin{equation}
T_{ba}=\frac{\sum_{l=1}^{n_b}\delta_{i(l)a}w(l)t(l)}{\sum_{l=1}^{n_b}w(l)},
\label{T simu}
\end{equation}
respectively. 
Here, a total of $n_b$ trajectories are initiated from milestone $b$, $i(l)$ is the index of the termination milestone of the $l$-th trajectory, $\delta_{i(l)a}$ is the Kronecker delta function, and $w(l)$ and $t(l)$ are the weight and time duration of the $l$-th trajectory, respectively.
Readers are referred to Ref. \dcite{LPTM} for more detailed discussions about the LPT-M method.

For latter kinetic analysis, it is more convenient to work in the Laplace space. As such, Eq. \eqref{basic eq} is transformed into
\begin{equation}
\mathbf{\tilde{q}}^T(z)=\mathbf{p}^T(0)[\mathbf{I}-\mathbf{\tilde{K}}(z)]^{-1},
\label{q eq in LaS}
\end{equation} 
where $\mathbf{\tilde{q}}(z)$ is a column vector with $\tilde{q}_a(z)=\int_0^\infty e^{-zt}q_a(t) dt$, $\mathbf{I}$ is an $N\times N$ identity matrix, and $\tilde{K}_{ba}(z)=\int_0^\infty e^{-zt}K_{ba}(t) dt$.

\subsection{Clustering}\label{Clustering}
As the number of milestones $N$ increases, the complexity of the kinetic network also grows, making it progressively challenging to be intuitively understood.
It is thus desirable to reduce the kinetic network by retaining only those long-lived metastable states.

The stochastic transition process in the discretized milestone space is generally non-Markovian. 
To facilitate clustering, an auxiliary CTMC is constructed and is described by a transition rate matrix $\mathbf{Q}$, where $Q_{ba}=K_{ba}/t_b$ for $a\neq b$ and $Q_{bb}=-\sum_{a\neq b}Q_{ba}$. 
The stationary probability $(\pi_a)_{a\in\mathcal{M}}$ of being at each milestone can be obtained by solving the eigenequation,
\begin{equation}
\mathbf{\pi}^T\mathbf{Q}=\mathbf{0}^T.
\label{equi prob}
\end{equation}
It can be readily checked that both the auxiliary CTMC and the stochastic transitions in Milestoning share the same stationary probabilities, $\pi_a=q_at_a$ ($q_a=\lim_{t\rightarrow\infty} q_a(t)$ is the stationary flux), $\forall a\in \mathcal{M}$, local residence time and MFPT between any two milestones\cite{Mile19}, providing a basis for locating metastable states.

For a time-reversible CTMC, the eigenvalues of $\mathbf{Q}$ are real-valued, arranged in descending order as $\mu_1=0\geq\mu_2\geq\cdots\geq\mu_N$.
The magnitude of an eigenvalue determines the decay rate of the corresponding mode. 
In the ideal case of a system composed of $k$ uncoupled subsystems, the multiplicity of eigenvalue $0$ is of $k$, i.e., $\mu_1=\mu_2=\cdots=\mu_k=0$.  
The corresponding $k$ right eigenvectors represent characteristic functions of invariant sets, denoted by $\mathbf{v}^T_i=(0,\cdots,0,\underbrace{1,\cdots,1}_{N_i},0,\cdots,0)$ (up to a permutation), with $i=1,2,\cdots,k$ and $\sum_{i=1}^k N_i=N$.

For an ergodic system composed of $k$ weakly coupled subsystems, the multiplicity of eigenvalue $0$ is one ($\mu_1=0$).
A spectral gap exists between the $k$-th and $(k+1)$-th eigenvalues ($\mu_k$ and $\mu_{k+1}$), indicating a timescale separation.
The first $k$ right eigenvectors contain characteristic information of almost invariant sets (metastable states).
In the PCCA+ algorithm, these $k$ eigenvectors are linearly combined to assign the most probable subsystem for each state\cite{PCCA}.

However, in networks generated from practical simulations, the time-reversibility may not be strictly preserved due to statistical fluctuations. 
As a result, the eigenvalues and eigenvectors of $\mathbf{Q}$ may become complex-valued, which makes it difficult to cluster in the eigenvector space.  

Here, we circumvent this problem by introducing a composite matrix $\mathbf{A}=\mathbf{Q}\tilde{\mathbf{Q}}$. 
The time-reversed transition rate matrix $\tilde{\mathbf{Q}}$ is defined as,
\begin{equation}
\tilde{\mathbf{Q}}=\boldsymbol{\Pi}^{-1}\mathbf{Q}^T\boldsymbol{\Pi},
\label{q tilde}
\end{equation}
where $\boldsymbol{\Pi}$ is a diagonal matrix with diagonal elements being the stationary probabilities,
\begin{equation}
\boldsymbol{\Pi} = \begin{pmatrix}
\pi_1 & 0 & \cdots & \cdots & 0\\
0 & \pi_2 & 0 & \cdots & 0\\
0 & 0 & \pi_3 & \cdots & 0\\
\vdots & \vdots & \vdots & \ddots &\vdots\\
0 & \cdots & \cdots & 0 & \pi_N\\
\end{pmatrix}.
\end{equation}

The matrix $\mathbf{A}$ has several nice properties (see Appendix \ref{app1} for the proof): (i) the matrix $\mathbf{A}$ satisfies the detailed balance condition, $\pi_aA_{ab}=\pi_bA_{ba}$. 
(ii) Its eigenvalues are real and non-negative, $0=\lambda_1\leq\lambda_2\leq\cdots\leq\lambda_N$. 
In particular for a time-reversible $\mathbf{Q}$, we have $\tilde{\mathbf{Q}}=\mathbf{Q}$. 
Consequently, the eigenvalues of $\mathbf{A}$ are related to those of $\mathbf{Q}$ via $\lambda_i=\mu_i^2$. 
(iii) In the ideal case of $k$ uncoupled subsystems ($\lambda_1=\cdots=\lambda_k=0$), the first $k$ right eigenvectors represent characteristic functions of invariant sets, i.e., $\mathbf{u}^T_i=(0,\cdots,0,\underbrace{1,\cdots,1}_{N_i},0,\cdots,0)$ up to a permutation, with $i=1,2,\cdots,k$ and $\sum_{i=1}^k N_i=N$.

The deviation from time-reversibility in $\mathbf{Q}$ due to statistical fluctuations is expected to be small, such that spectral structure still exists, especially for those eigenvalues close to zero. 
A spectral gap is assumed to be in between $\lambda_k$ and $\lambda_{k+1}$.
Hence, the first $k$ right eigenvectors of $\mathbf{A}$ are used to classify all milestones into $k$ clusters.
In particular, the $k$-means algorithm is utilized to assign each milestone to a specific cluster in the space spanned by $\{\mathbf{u}_2,\mathbf{u}_3,\cdots,\mathbf{u}_k\}$, where $\mathbf{u}_1^T=(1,\cdots,1)$ has been trivially ruled out because its elements are identical for each milestone. 

The location of spectral gaps is usually self-evident when there is an obvious timescale separation and is empirically determined in later numerical tests.

\subsection{Kinetic Network Reduction}\label{Reduction}
After clustering, the next step is to identify the core set of each cluster. 
The ultimate goal is to construct a reduced kinetic network consisting of only these core sets as nodes.
Here, the core set $C_i$ of the $i$-th cluster $D_i$ is defined as the milestone with the lowest free energy in $D_i$, $c^i_{min}={arg\,min}_{a\in D_i} F_a$, plus those directly connected to it within a certain free energy cutoff.
Mathematically, this is expressed as:
\begin{equation}
C_i=\{c^i_{min}\}\cup\{a\in D_i:|F_a-F_{c^i_{min}}|<F_{cut},\ \max(K_{ac^i_{min}},K_{c^i_{min}a})>0\}.
\label{core sets criterion}
\end{equation}
Two milestones $a$ and $b$ are considered connected if either $K_{ab}$ or $K_{ba}$ is nonzero.
The free energy of a milestone is related to its stationary probability, $F_a=-k_BT\ln\pi_a$. 
The free energy cutoff $F_{cut}$ is a parameter and is set to be $1.5\ k_BT$ in the current study.

\begin{figure}[h]
\centering
\begin{tabular}{cc}
\includegraphics[height=5cm]{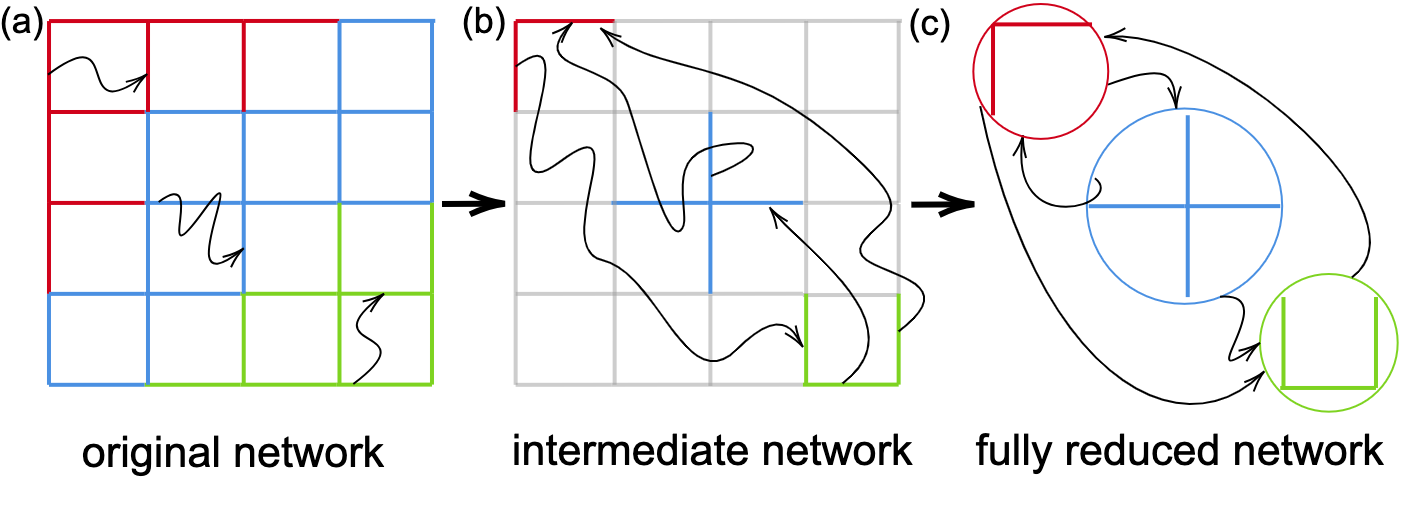}\\
\end{tabular}
\caption{Schematic illustration of a two-step reduction of the kinetic network in Milestoning. (a) In this example, the original network contains $40$ milestones, which are grouped into three clusters (coded in different colors). The dimensions of $\mathbf{K}$ and $\mathbf{T}$ (calculated by Eqs. \eqref{K simu} and \eqref{T simu}, respectively) are both $40\times 40$. (b) Three core sets (colored in red, blue and green, respectively) are identified within clusters, and all other milestones states (in gray) are removed from the original network. Milestones in these three core sets constitute the intermediate network. In the intermediate network, only transitions among milestones in different core sets are considered. The dimensions of $\mathbf{K}^{(I)}$ and $\mathbf{T}^{(I)}$ (calculated by Eqs. \eqref{K inter} and \eqref{T inter}, respectively) are both $9\times 9$ here. (c) Each core set is further grouped as a single whole state, resulting in a fully reduced network. The core sets in red and green are designated as the reactant and product state, respectively. The dimensions of $\mathbf{K}^{(R)}$ and $\mathbf{T}^{(R)}$ (calculated by Eqs. \eqref{K reduced} and \eqref{T reduced}, respectively) are both $3\times 3$ here.}\label{fig_reduction}
\end{figure}

Now, we proceed to calculate the transition probabilities and transition time among these core sets, which is performed in two steps (cf. Fig. \ref{fig_reduction}). 

\textit{Step I. Original network $\rightarrow$ Intermediate network.} 
In the intermediate network, milestones not belonging to core sets are removed from the original network.
The remaining milestones constitute a new set, denoted as $\mathcal{C}=\{\underbrace{a,\cdots,b}_{C_1},\cdots,\underbrace{g,\cdots,h}_{C_k}\}$.
Additionally, milestones within the same core set are made invisible to each other.
Subsequently, the transition probabilities and transition time among these remaining milestones are recalculated. 

The new transition probability $K^{(I)}_{df}$ ($d$ and $f$ belong to different core sets) amounts to the splitting probability of $d\rightarrow f$ and can be calculated by
\begin{equation}
K^{(I)}_{df}=\int_0^\infty q_f(t)dt=\tilde{q}_f(0),
\label{KI def}
\end{equation}
where the superscript $I$ indicates the intermediate network, and the initial condition is set on milestone $d$. 
Using Eq. \eqref{q eq in LaS}, the splitting probability of $d\rightarrow f$ is given by
\begin{equation}
K^{(I)}_{df}=\mathbf{e}_d^T(\mathbf{I}-\mathbf{K}')^{-1}\mathbf{e}_f,
\label{K inter}
\end{equation}
where $\mathbf{e}_d$ and $\mathbf{e}_f$ are two unit column vectors with only the $d$-th and $f$-th elements being one, respectively, and the other elements being zero. 
Here, $\mathbf{K}'$ denotes a modified transition probability matrix of the original network with absorbing boundary conditions imposed at all milestones in $\mathcal{C}$ except the core set where $d$ is located, i.e.,
\begin{equation}
K'_{ba}=\left\{
    \begin{array}{llllll}
    0, & b\in \mathcal{C}\backslash C_d\ (\mathrm{assume}\ d\in C_d),\\
    K_{ba}, & \mathrm{otherwise}.
    \end{array}
\right.
\end{equation} 
Equation \eqref{K inter} is a natural generalization of the committor expression\cite{CommMile,ComminMile}, where only two end states are considered.

The transition time from milestone $d$ to $f$ is given by
\begin{equation}
T^{(I)}_{df}=\frac{\int_0^\infty tq_f(t)dt}{K^{(I)}_{df}}=\frac{-\frac{d}{dz}\tilde{q}_f(z)|_{z=0}}{K^{(I)}_{df}}.
\label{TI def}
\end{equation} 
Using Eq. \eqref{q eq in LaS}, the transition time of $d\rightarrow f$ becomes
\begin{equation}
T^{(I)}_{df}=\frac{\mathbf{e}^T_d(\mathbf{I}-\mathbf{K}')^{-1}\mathbf{T}'(\mathbf{I}-\mathbf{K}')^{-1}\mathbf{e}_f}{K^{(I)}_{df}},
\label{T inter}
\end{equation}
where $\mathbf{T}'=\int_0^\infty t\mathbf{K}'(t)dt$.
Equation \eqref{T inter} is a conceptual generalization of the exit time to product expression derived in Ref. \dcite{TranTime}.

Given Eq. \eqref{T inter}, the average residence time on milestone $d$ in the intermediate network can be calculated by
\begin{equation}
t_d^{(I)} = \sum_{f\in\mathcal{C}}T_{df}^{(I)}K_{df}^{(I)} = \mathbf{e}_d^T(\mathbf{I}-\mathbf{K}')^{-1}\mathbf{t}',
\label{t inter}
\end{equation}
with $\mathbf{t'}=\mathbf{T}'\mathbf{1}_N$. 
That is, the average residence time can be directly evaluated without first calculating $T_{df}^{(I)}$ explicitly.

It can be checked that with the transition probability and average residence time calculated via Eqs. \eqref{K inter} and \eqref{t inter}, respectively, the MFPT between any two milestones in different core sets remains unchanged before and after the reduction (see Appendix \ref{app2} for the sketch of the proof).

\textit{Step II. Intermediate network $\rightarrow$ Fully reduced network.} Milestones in each core set are further combined into a single node to arrive at a fully reduced network.  
The goal is to ensure that the fully reduced network retains the same MFPT from the reactant state (designated as the core set $C_i$) to the product state (designated as the core set $C_f$).
Thus, the combinational coefficients must be appropriately chosen.

To screen out the unidirectional transitions of $C_i\rightarrow C_f$ and obtain the stationary fluxes, we solve the eigenequation,
\begin{equation}
\mathbf{q}^T=\mathbf{q}^T\mathbf{K}^{(I)'},
\label{q stationary}
\end{equation}
where $\mathbf{q}$ is a column vector of length $|\mathcal{C}|$ with each element being the stationary flux through each milestone in core sets $\mathcal{C}$, and $\mathbf{K}^{(I)'}$ is a modified transition probability matrix of the intermediate network with cyclic boundary conditions imposed in $C_f$, i.e.,
\begin{equation}
K^{(I)'}_{ba}=\left\{
    \begin{array}{llllll}
    p_a(0), & b\in C_f\ \mathrm{and}\ a\in C_i,\\
    0, & b\in C_f\ \mathrm{and}\ a\notin C_i,\\
    K^{(I)}_{ba}, & \mathrm{otherwise}.
    \end{array}
\right.
\end{equation}
Equation \eqref{q stationary} is a reformulation of Eq. \eqref{basic eq} in the stationary limit $t\rightarrow\infty$, but is now applied to the intermediate network. 

The stationary fluxes serve as the weighting factors to combine milestones in each core set.
The transition probability and transition time from the core set $C_a$ to $C_b$ in the fully reduced network are thus given by
\begin{equation}
K^{(R)}_{C_aC_b}=\frac{\sum_{a\in C_a}q_a\sum_{b\in C_b}K^{(I)'}_{ab}}{\sum_{a\in C_a}q_a},
\label{K reduced}
\end{equation}
and
\begin{equation} 
T^{(R)}_{C_aC_b}=\frac{\sum_{a\in C_a}q_a[\sum_{b\in C_b}T^{(I)'}_{ab}K^{(I)'}_{ab}]}{\sum_{a\in C_a}q_a}\cdot\frac{1}{K_{C_aC_b}^{(R)}},
\label{T reduced}
\end{equation}
respectively, where $\mathbf{T}^{(I)'}$ is different from $\mathbf{T}^{(I)}$ in boundary conditions at $C_f$,
\begin{equation}
T^{(I)'}_{ab}=\left\{
    \begin{array}{llllll}
    0, & a\in C_f,\\
    T^{(I)}_{ab}, & \mathrm{otherwise}.
    \end{array}
\right.
\end{equation}
The term inside the bracket in the numerator of Eq. \eqref{T reduced} is the unnormalized transition time from a milestone $a$ in $C_a$ to the core set $C_b$.
Based on Eqs. \eqref{K reduced} and \eqref{T reduced}, the average residence time on $C_a$ can also be evaluated,
\begin{equation}
t^{(R)}_{C_a} =\sum_{b=1}^kT^{(R)}_{C_aC_b}K^{(R)}_{C_aC_b} = \frac{\sum_{a\in C_a}q_at_a^{(I)'}}{\sum_{a\in C_a}q_a}.
\label{t reduced}
\end{equation}
Again, $\mathbf{t}^{(I)'}$ is different from $\mathbf{t}^{(I)}$ only at $C_f$, i.e., $t_{f}^{(I)'}=0, \forall f\in C_f$.

\subsection{Transition Path Analysis}\label{Pathway}
To facilitate understanding the mechanism of the transition process from the reactant state $C_i$ to the product state $C_f$, transition paths can be analyzed on the reduced Milestoning network using the graph theory.
A graph is denoted as $G(S,E)$, where the core sets $\{C_1,\cdots,C_k\}$ constitute the nodes $S$ of the graph. 
In the same spirit of TPT\cite{TPTMJ}, the effective current between two nodes is defined as the directed edge $E$. 
The edge weight is calculated as 
\begin{equation}
E_{C_aC_b}=|q^{(R)}_{C_a}K^{(R)}_{C_aC_b}-q^{(R)}_{C_b}K^{(R)}_{C_bC_a}|,
\label{edge weight}
\end{equation}
and its direction is from $C_a$ to $C_b$ if $q^{(R)}_{C_a}K^{(R)}_{C_aC_b} > q^{(R)}_{C_b}K^{(R)}_{C_bC_a}$ and vice versa.
Here, $\mathbf{q}^{(R)}$ is the stationary flux vector through core sets and is obtained by solving the eigenequation
\begin{equation}
(\mathbf{q}^{(R)})^T\mathbf{K}^{(R)}=(\mathbf{q}^{(R)})^T.
\end{equation}  

A transition path of $C_i\rightarrow C_f$ is denoted as $\omega=(C_i,\cdots
,C_a,C_b,\cdots,C_f)$.
In the transition path ensemble $\Omega$, transition paths with maximum min-weight are of particular interest.
The min-weight of a path is defined as
\begin{equation}
E_{min}(\omega)=\min_{(C_a,C_b)\in\omega}E_{C_aC_b}.
\end{equation}
The maximum min-weight path (MWP) has the largest min-weight (bottleneck) edge in all transition paths connecting $C_i$ and $C_f$,
\begin{equation}
\omega_{MWP}=\{\omega\in\Omega: E_{min}(\omega)=\max_{\omega'\in\Omega}\{E_{min}(\omega')\}\}.
\end{equation} 
The edge weight is also interpreted as the capacity in computer science\cite{CSCap60,CSCap61}.
Thus, MWPs have the maximum capacity in transporting trajectories from the reactant to the product state and are referred to as dominant reaction pathways in TPT\cite{TPTMJ}.

However, MWPs may not be unique as different paths may share the same bottleneck edge.
Therefore, it is useful to consider the transition path with global maximum min-weight, which is as close as possible to being unique.
The global maximum min-weight path (GMWP) is defined as follows: for any two nodes in GMWP, the corresponding segment is a MWP among all possible paths connecting these two nodes.
The GMWP is referred as the representative dominant reaction pathway in TPT\cite{TPTMJ} and has also been investigated in the Milestoning network in Ref. \dcite{MileNet}. 
But the discussions here are different from Ref. \dcite{MileNet} in two aspects: (i) the effective current is defined as the edge weight in this paper, which is in the same spirit of TPT. (ii) The pathway analysis is performed in the reduced Milestoning network, providing a more intuitive picture of the transition process.

The recursive Dijkstra's algorithm is used to find the GMWP\cite{MileNet}.
Once the GMWP is identified, we may proceed to find the next transition path with the global second largest min-weight edge.
To achieve this, the edge weights first need to be adjusted\cite{TPTMJ},
\begin{equation}
E'_{C_aC_b}=\left\{
    \begin{array}{llllll}
    E_{C_aC_b}-E_{min}(\omega_{GMWP}), & \mathrm{if}\ (C_a,C_b)\in\omega_{GMWP},\\
    E_{C_aC_b}, & \mathrm{otherwise}.
    \end{array}
\right.
\label{edge weight adjust}
\end{equation} 
That is, the edge weights along GMWP are updated by subtracting the bottleneck edge weight, while the other edge weights remain unchanged.
Then the recursive Dijkstra's algorithm is invoked again to find a GMWP in the updated graph induced by $G(S,E')$.
This process can be repeated until the reactant state and the product state are disconnected.

If there is no path degeneracy, the bottleneck edge weights of GMWPs found in sequence are ordered as $E_{min}(\omega^{(1)}_{GMWP})>E_{min}(\omega^{(2)}_{GMWP})>\cdots$, where the superscript indicates the order index.
These bottleneck edge weights are used to calculate the contribution ratio of each path,
\begin{equation}
R(\omega^{(i)}_{GMWP})=\frac{E_{min}(\omega^{(i)}_{GMWP})}{\sum_j E_{min}(\omega^{(j)}_{GMWP})}.
\label{path ratio}
\end{equation}

\section{Computational Details}\label{Simulation}
\subsection{Two-dimensional Model}
The potential energy surface of the two-dimensional model system $U(x,y)$ has the following form,
\begin{align}
U(x,y) =& 3\exp[-x^2-(y-0.2)^2]-3\exp[-x^2-(y-1.8)^2]\nonumber\\
&-5\exp[-y^2-(x-1.0)^2]-5\exp[-y^2-(x+1.0)^2]\nonumber\\
&+10^{[x^2+(y-0.5)^2-9]}.
\label{model_pes}
\end{align} 
Overdamped Langevin dynamics is evolved on the energy landscape following the equation
\begin{equation}
\dot{\mathbf{r}}=-\nabla U(x,y)+\mathbf{\eta}.
\end{equation}
The Euler-Maruyama algorithm is utilized with the integration time step $\Delta t=10^{-4}$ and temperature $k_BT=1$. 
The white noise $\mathbf{\eta}$ is of mean zero and covariance $\langle\mathbf{\eta}_i(t)\mathbf{\eta}_j(t')\rangle=2k_BT\delta(t-t')\delta_{ij}$.

The configuration space is randomly partitioned into 56 cells using the Voronoi tessellation, resulting in a total number of 141 milestones (Fig. \ref{fig_model_spec} (a)). 
400 trajectories are initiated from each milestone for LPT-M analysis. 

\subsection{Solvated Alanine Dipeptide}
The alanine dipeptide is modeled as ACE-ALA-NME (Fig. \ref{fig_ala2_spec} (a)), where the peptide termini are capped with an acetyl group (ACE) and an N-methyl amide group (NME), respectively.
The program NAMD 2.14\cite{NAMD} and CHARMM36 force fields\cite{CHARMM36} are used for all simulations.
The system is solvated with 742 TIP3P water molecules.
Periodic boundary conditions are applied in all three directions.
The system is minimized with a conjugate gradient algorithm for 10000 steps and then equilibrated in the NPT ensemble using the Nose-Hoover Langevin piston pressure control\cite{NPT94,NPT95} for 3 ns at a pressure of 1 atm and a temperature of 300 K. 
Finally, the system is equilibrated in the NVT ensemble for 2 ns at 300 K using a Langevin thermostat before the Milestoning calculation.

In all simulations, water molecules are kept rigid using the SETTLE algorithm\cite{SETTLE}, and all other bond lengths with hydrogen atoms are kept fixed using the SHAKE algorithm\cite{SHAKE}.
The integration time step is 1 fs.
A real space cutoff distance of 9\r{A} is used for both electrostatic and van der Waals interactions, and particle mesh Ewald is utilized for long-range electrostatic calculations\cite{PME}. 

The Milestoning calculations are performed with two dihedral angels ($\phi$ and $\psi$) and four dihedral angles ($\phi$, $\psi$, $\theta$ and $\zeta$), respectively.
The definition of these involved dihedral angels is illustrated in Fig. \ref{fig_ala2_spec} (a).

In the case of the two-angle calculations, the phase space is uniformly divided into 64 cells with an equal interval of 45\degree\ in $\phi$ and $\psi$, respectively.
This partition leads to a total number of 128 milestones.
400 trajectories are initiated from each milestone and are analyzed using the LPT-M method.

In the case of the four-angle calculations, the MD simulations keep the amide planes in trans configuration, so $\theta$ and $\zeta$ are restrained in the range $[-60\degree, 60\degree]$.
Each milestone defined in the $(\phi,\psi)$ space in the two-angle calculation is further split into four milestones.
That is, for a given milestone in $(\phi, \psi)$, the orthogonal space in $(\theta, \zeta)$ is uniformly divided into four cells with an equal interval of $60\degree$ in each angle, $\{\theta\in[-60\degree, 0\degree]\cup [0\degree, 60\degree]\}\otimes\{\zeta\in[-60\degree, 0\degree]\cup [0\degree, 60\degree]\}$.
As a result, the total number of milestones is 512.
200 trajectories are initiated from each milestone and are analyzed using the LPT-M method.


\section{Results and Discussion}\label{Results}
\subsection{Two-dimensional Model}
\begin{figure}[h]
\centering
\begin{tabular}{cccc}
\makecell{
\includegraphics[height=5.5cm]{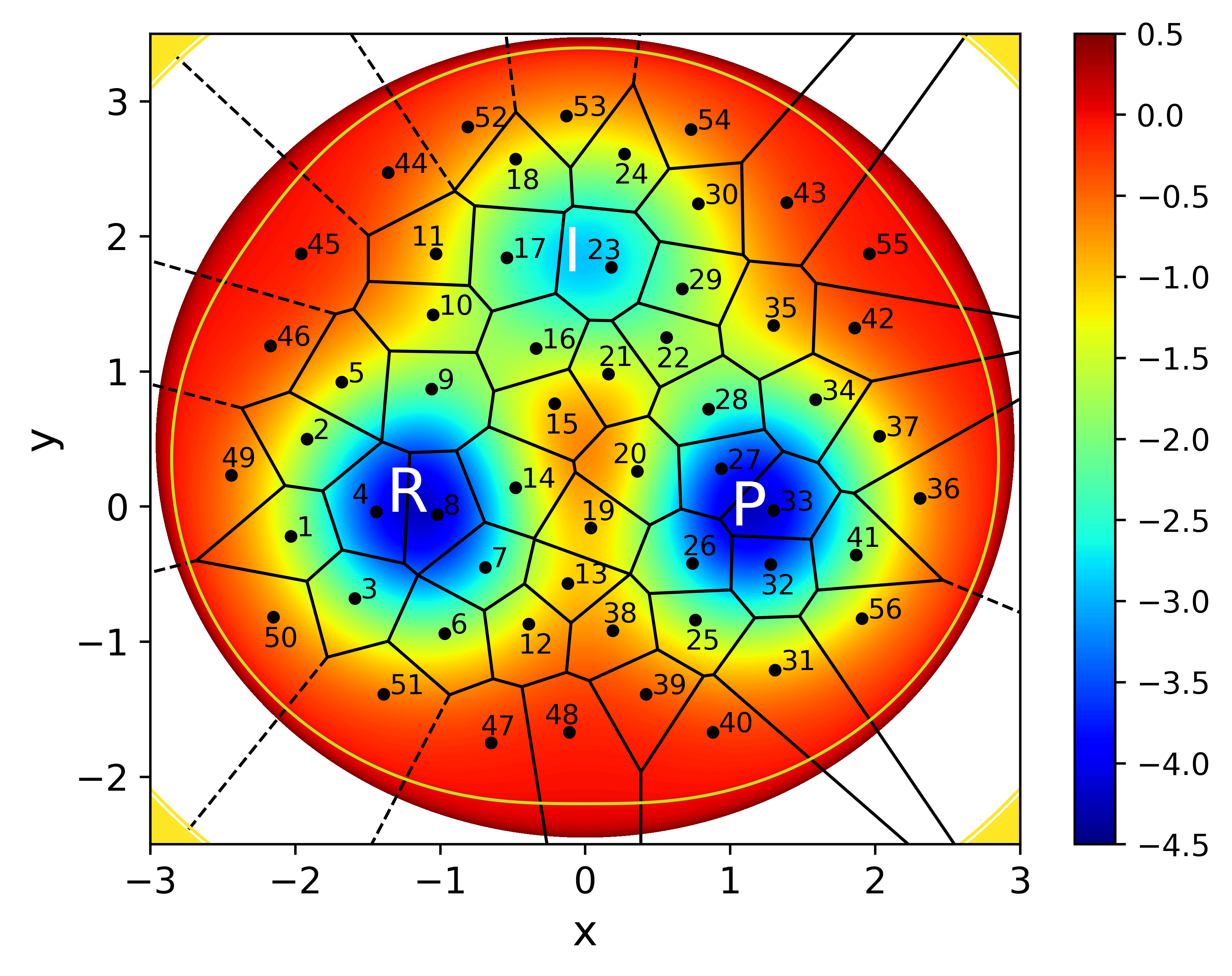}\\
(a)} \\
\makecell{
\includegraphics[height=5.5cm]{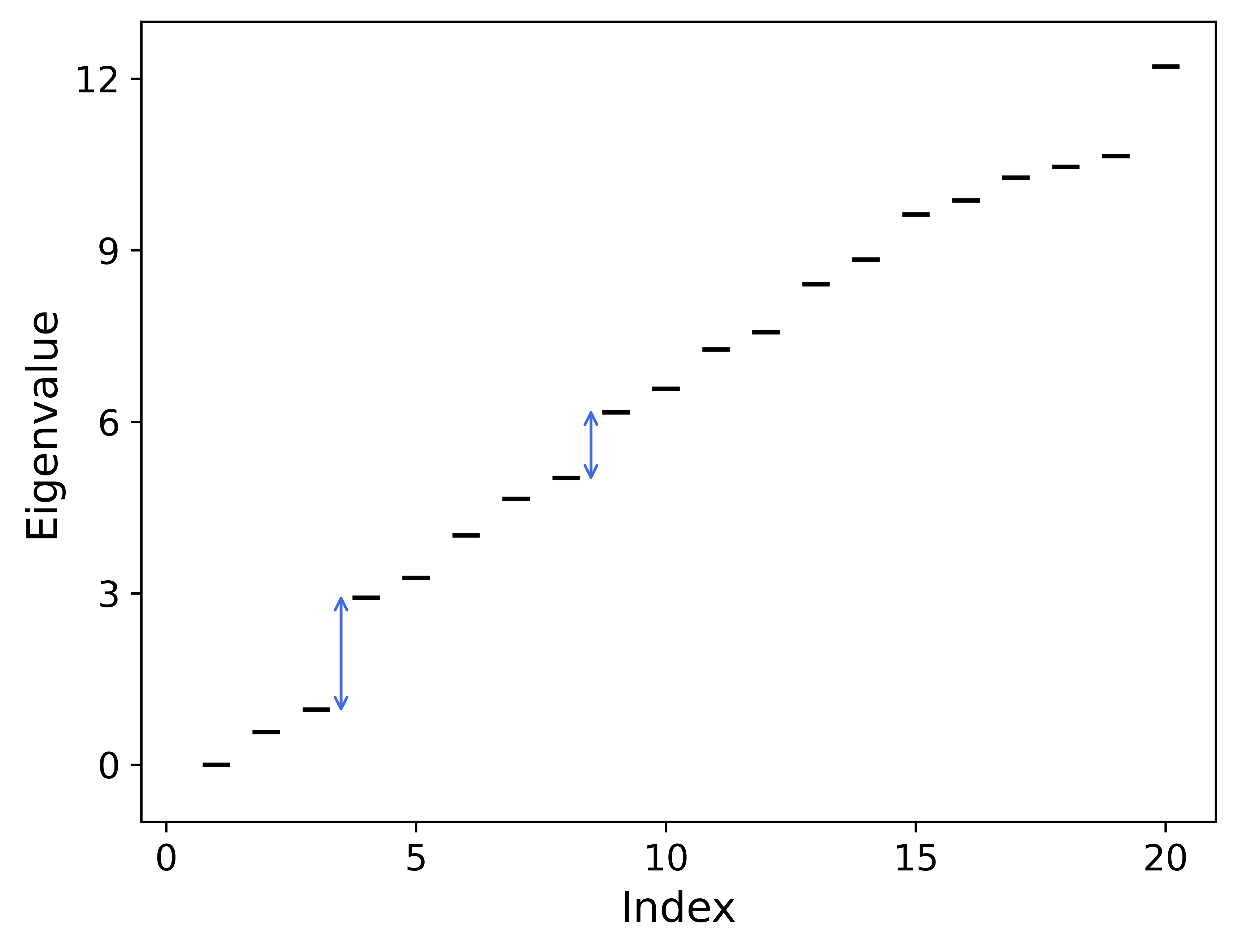} \\
(b)} \\
\makecell{
\includegraphics[height=5.5cm]{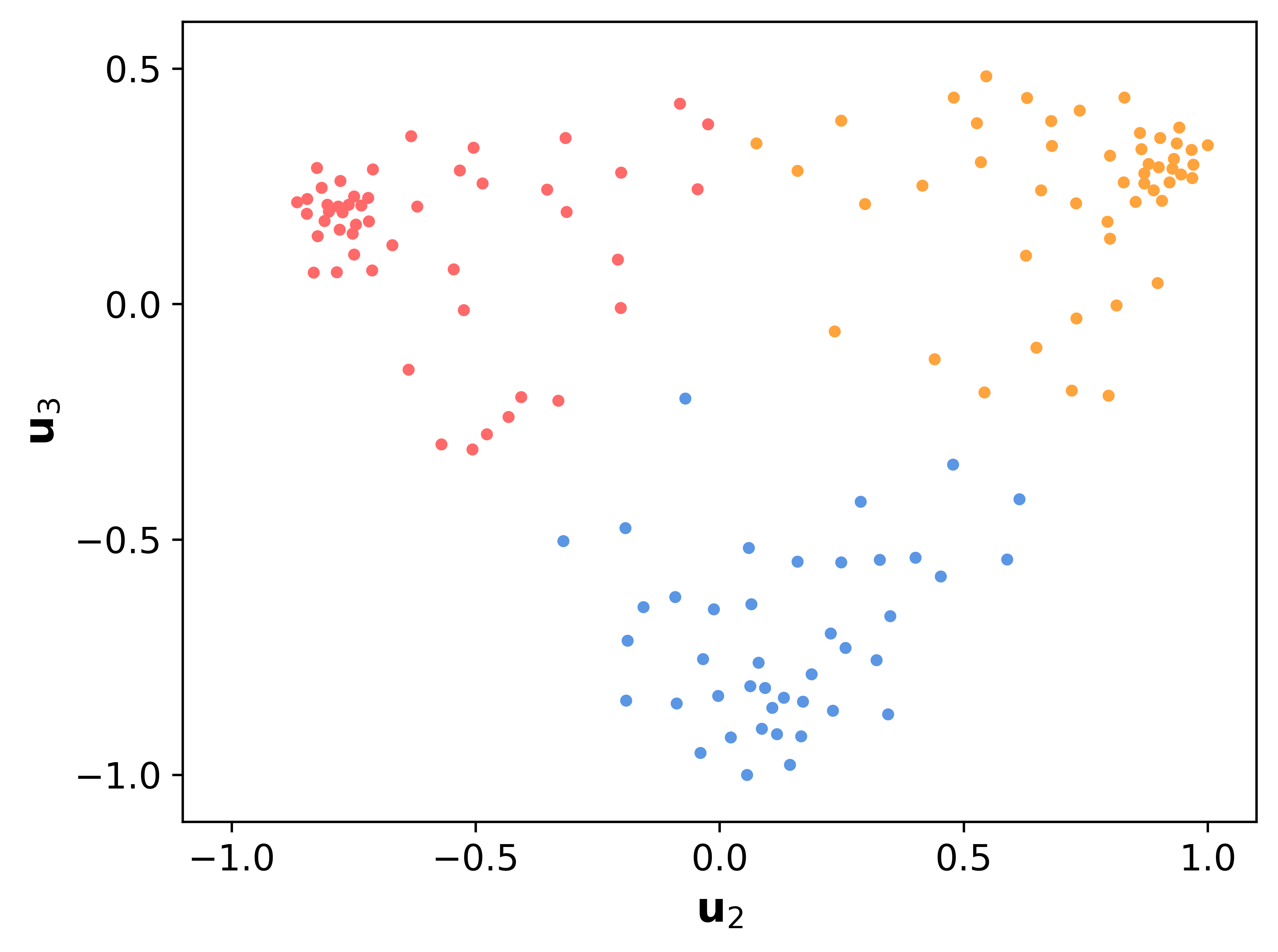} \\
(c)}\\
\end{tabular}
\caption{The two-dimensional model system. (a) The potential energy landscape (defined in Eq. \eqref{model_pes}) is partitioned into 56 cells. There are three minima: the reactant (R), the product (P) and an intermediate state (I). (b) The eigenvalues of the composite matrix $\mathbf{A}=\mathbf{Q}\tilde{\mathbf{Q}}$ in ascending order. The first two spectral gaps are indicated by arrows. (c) Clustering of milestones (dots) in the space spanned by the second and third right eigenvectors of $\mathbf{A}$.}\label{fig_model_spec}
\end{figure}

The energy landscape of the two-dimensional model has three minima (cf. Fig. \ref{fig_model_spec} (a)), with two designated as the reactant state (R) and the product state (P), respectively.
The third, shallower one represents an intermediate state (I).
This model system serves as a simple illustration of the features of molecular rearrangements.

\begin{figure}[h]
\centering
\begin{tabular}{cccccc}
\makecell{
\includegraphics[height=5cm]{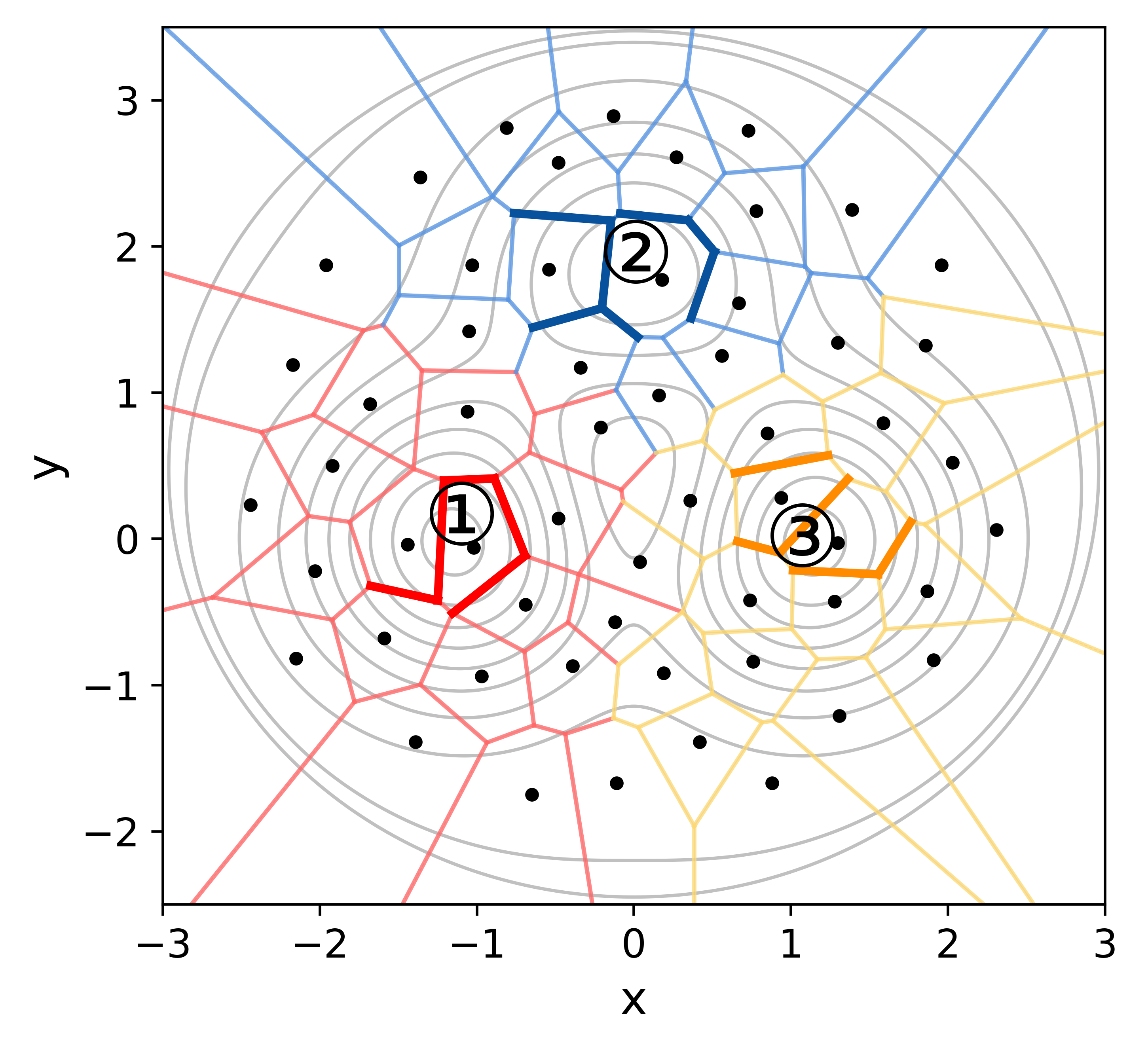} \\
(a)}
\makecell{
\includegraphics[height=5cm]{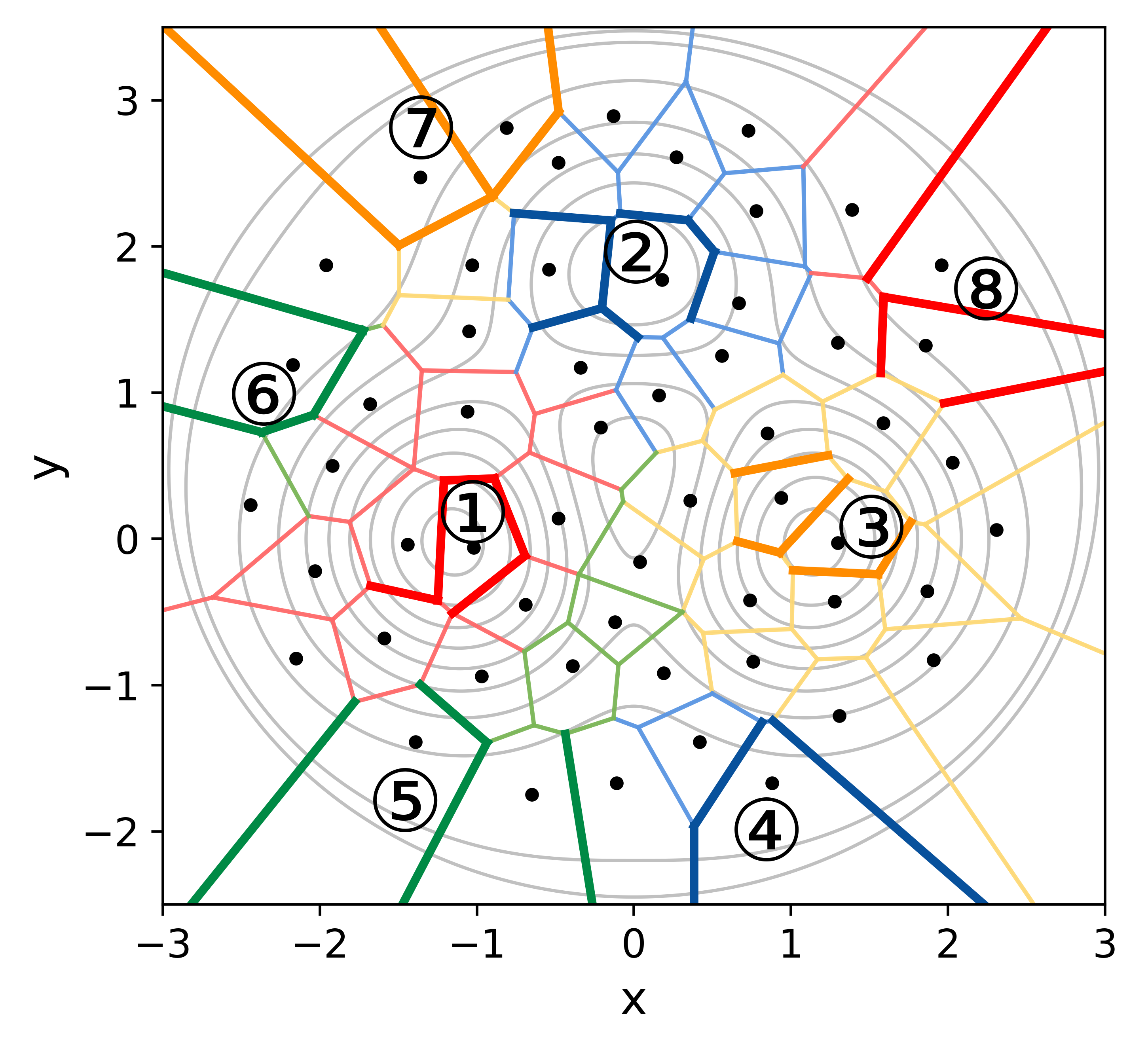} \\
(d)}\\
\makecell{
\includegraphics[height=5cm]{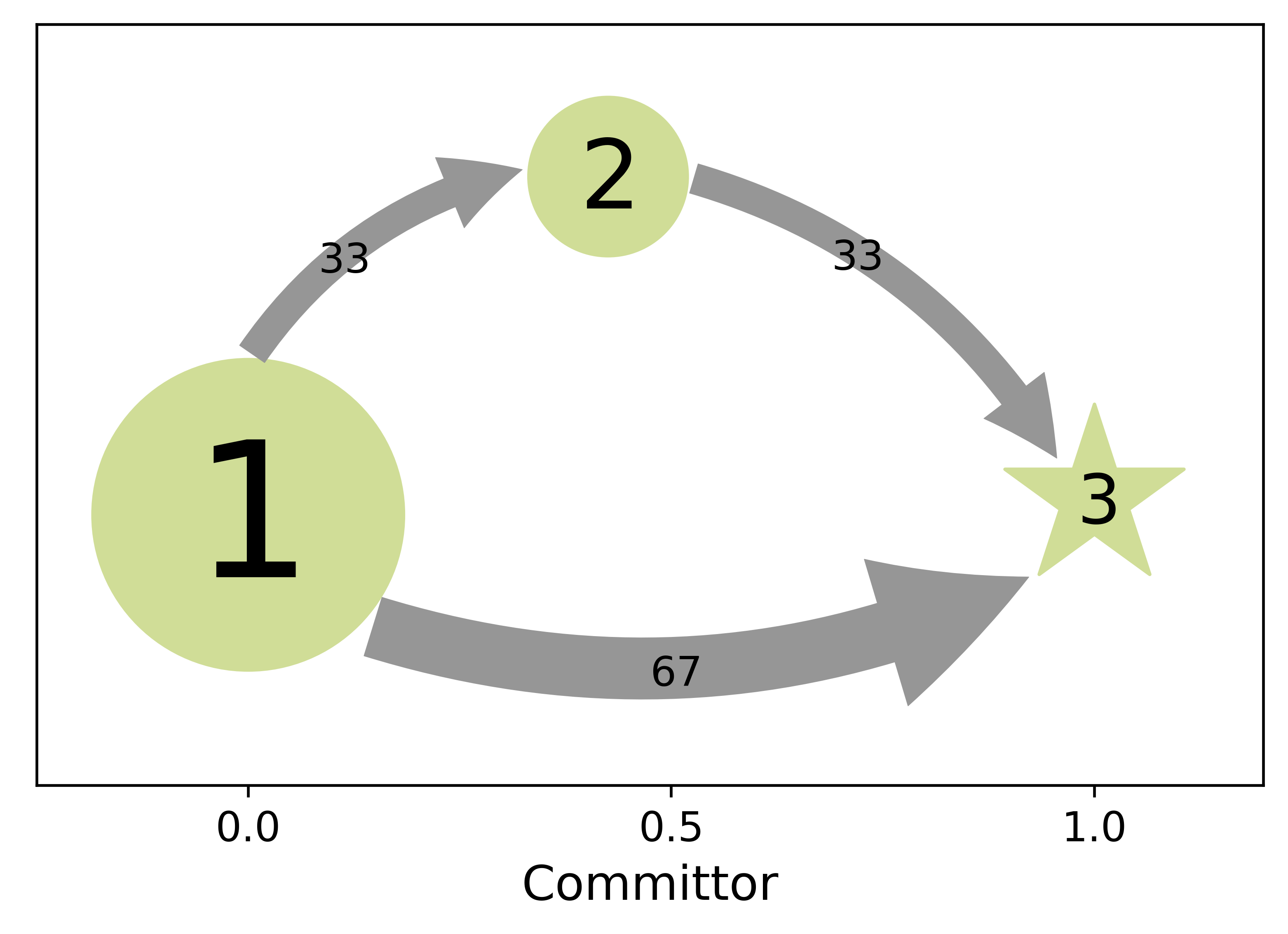} \\
(b)}
\makecell{ 
\includegraphics[height=5cm]{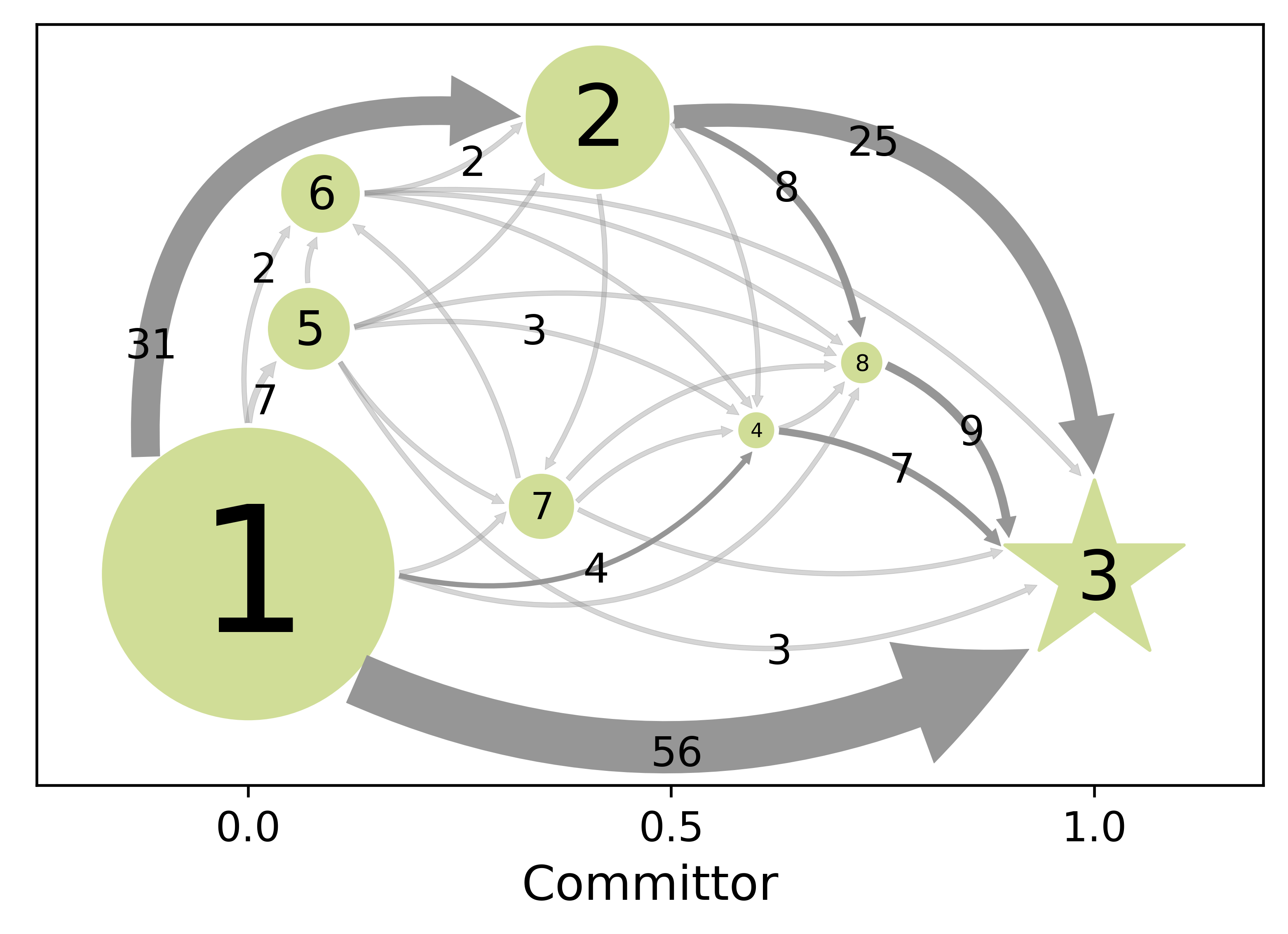} \\
(e)} \\
\makecell{
\includegraphics[height=4cm]{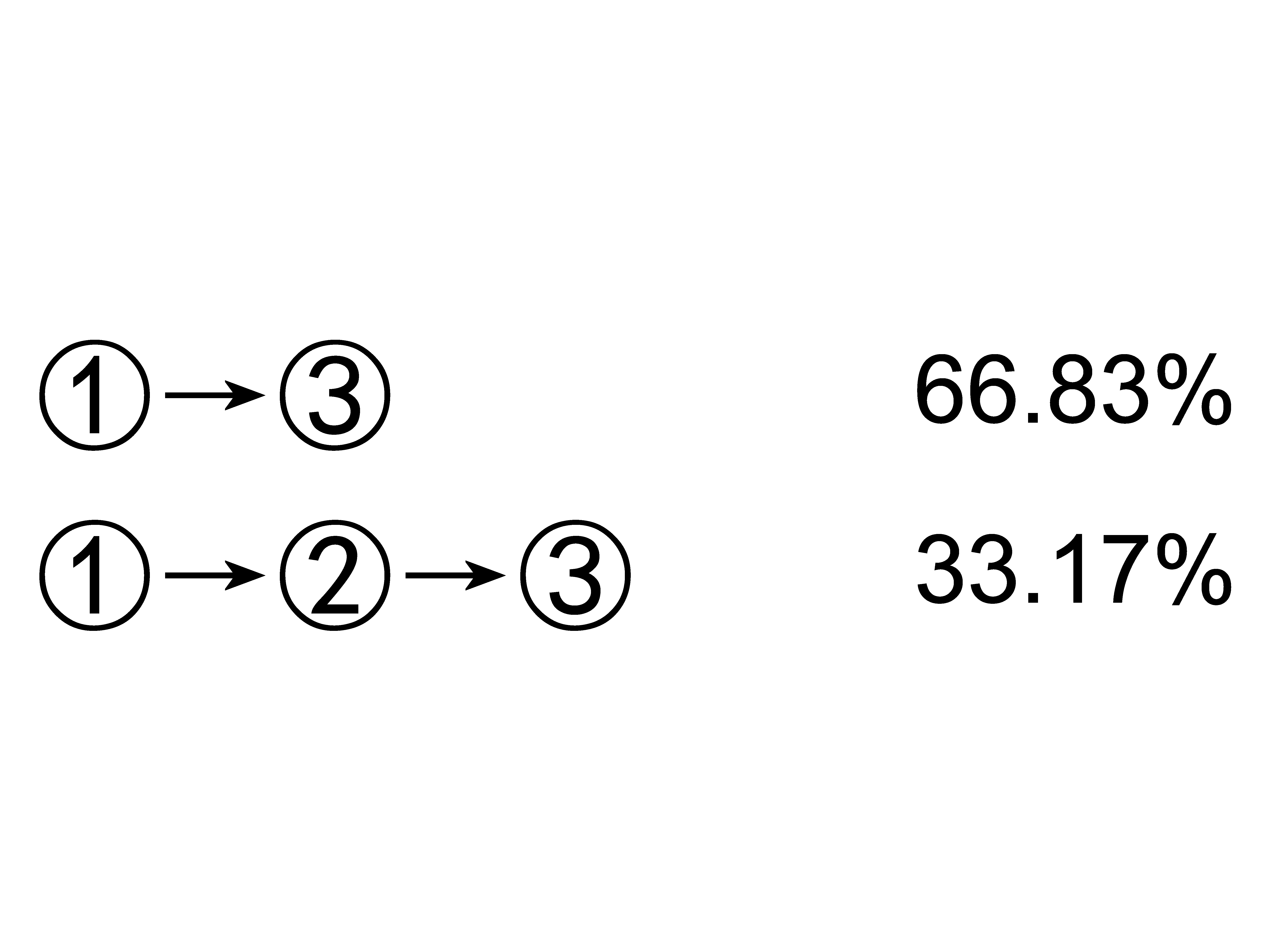}\\
(c)}
\makecell{
\includegraphics[height=4cm]{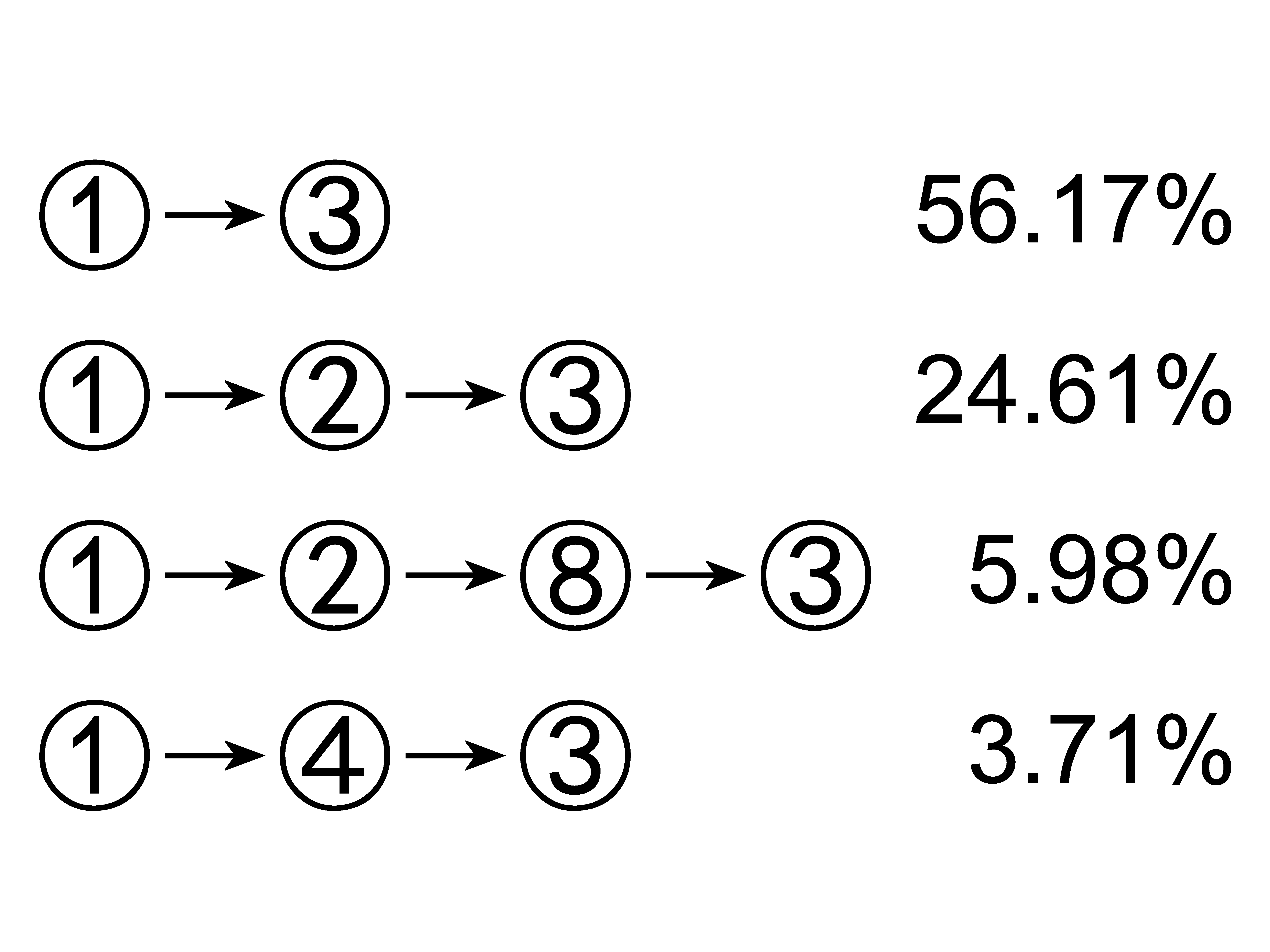} \\
(f)}\\
\end{tabular}
\caption{Clustering and reduction results for the two-dimensional model. The left panel ((a)-(c)) and right panel ((d)-(e)) illustrate the three-cluster and eight-cluster results, respectively. Top: Milestones are grouped into clusters (shown in different colors). The core set of each cluster is indicated by bold lines. Middle: The reduced network with normalized effective current (rounded off to integers). The total outflow current from the reactant state (\ding{172}) is set to 100. Effective currents smaller than 1 are not labeled. The size of each node (circle) is proportional to the stationary probability of the corresponding core set. The product state (\ding{174}) is marked as a star. Bottom: The first few dominant transition pathways and their contribution ratios are listed.}\label{fig_model_red}
\end{figure}

The spectral structure of the composite matrix $\mathbf{A}=\mathbf{Q}\tilde{\mathbf{Q}}$ is shown in Fig. \ref{fig_model_spec} (b).
The first spectral gap occurs between the third and fourth eigenvalues, indicating three main metastable states, which is consistent with the energy landscape.
Milestones are then grouped into three clusters in the space spanned by $(\mathbf{u}_2, \mathbf{u}_3)$ (Fig. \ref{fig_model_spec} (c)), in which three clusters are clearly separated.
The core set of each cluster is identified as described in Sec. \ref{Reduction} (cf. Fig. \ref{fig_model_red} (a)).
Transition probabilities and transition time among these three core sets are also calculated.
In the fully reduced network (Fig. \ref{fig_model_red} (b)), there are only three nodes, one for each core set.
Here, the transition probability of the intermediate state to the product state also represents the committor value.
For the three-cluster case, only two possible transition pathways exist (Fig. \ref{fig_model_red} (c)).
In the current simulation setup, the direct transition path is dominant, whose contribution is about twice as large as that of the indirect transition path.

The next spectral gap occurs between the eighth and ninth eigenvalues.
Milestones are then grouped into eight clusters (Fig. \ref{fig_model_red} (d)), with the newly added core sets having higher energy than the three main metastable states.
With more intermediate states being added, the resulting reduced network shows a more complex form (Fig. \ref{fig_model_red} (e)).
As such, transition pathway analysis will provide a more detailed description of the transition process.
As shown in Fig. \ref{fig_model_red} (f), the direct transition path (\ding{172}$\rightarrow$\ding{174}) and the indirect transition path via the intermediate state I (\ding{172}$\rightarrow$\ding{173}$\rightarrow$\ding{174}) are still the two dominant pathways, contributing over 80\%.
The remaining transition pathways need to pass through high-energy states and therefore contribute less.

\subsection{Solvated Alanine Dipeptide}
\subsubsection{Two-angle Calculation}
The alanine dipeptide serves as a classical test system for benchmarking new methods.
In the two-angle calculation, its conformational dynamics is characterized by two traditional backbone dihedral angles $\phi$ and $\psi$.
The free energy landscape in terms of $\phi$ and $\psi$ is calculated using the combined umbrella sampling method\cite{US74,US77} and the weighted histogram analysis method\cite{WHAM92,WHAM01} (WHAM) to be contrasted with the clustering result.

\begin{figure}[h]
\centering
\begin{tabular}{cccccc}
\makecell{
\includegraphics[height=5cm]{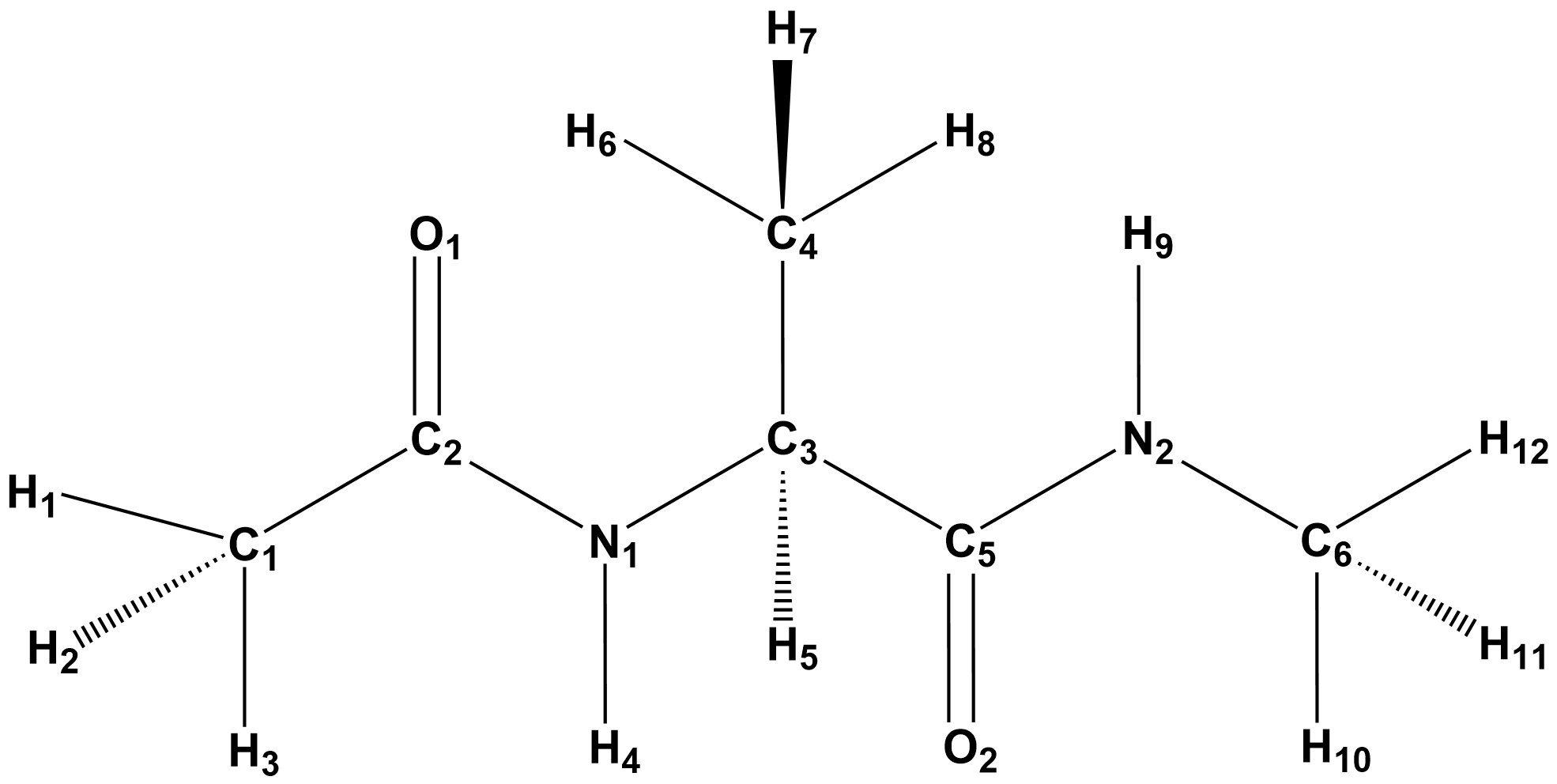} \\
(a)}\\
\makecell{
\includegraphics[height=6cm]{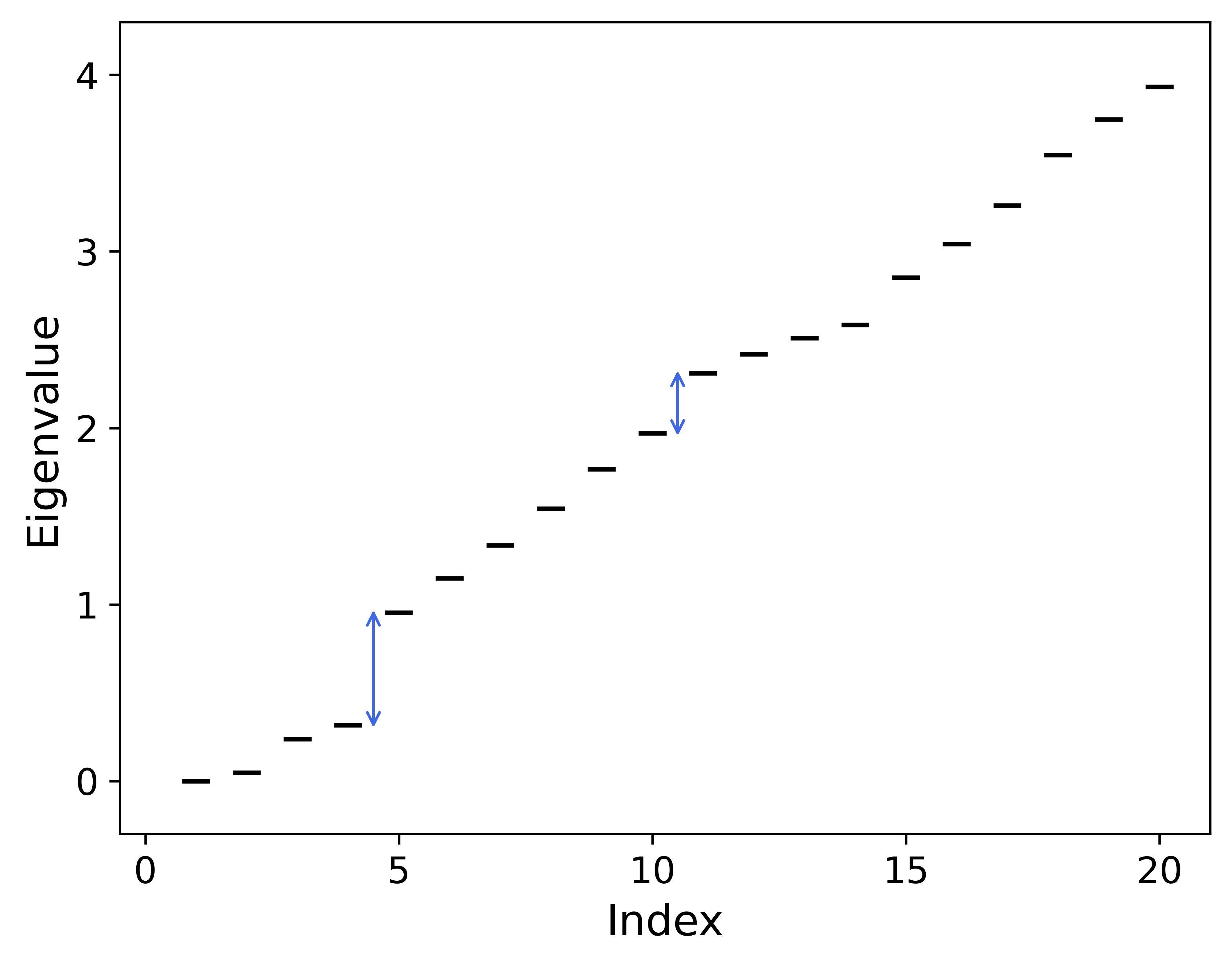} \\
(b)}\\
\end{tabular}
\caption{(a) Structure of the alanine dipeptide. Dihedral angles involved in Milestoning calculations are defined as follows: $\phi: C_2-N_1-C_3-C_5$, $\psi: N_1-C_3-C_5-N_2$, $\theta: O_1-C_2-N_1-C_3$, $\zeta: C_3-C_5-N_2-H_9$. (b) The eigenvalues of the composite matrix $\mathbf{A}=\mathbf{Q}\tilde{\mathbf{Q}}$ in ascending order for the two-angle calculation with the solvated alanine dipeptide. The first two spectral gaps are indicated by arrows.}\label{fig_ala2_spec}
\end{figure}

\begin{figure}[h]
\centering
\begin{tabular}{cccccc}
\makecell{
\includegraphics[height=5cm]{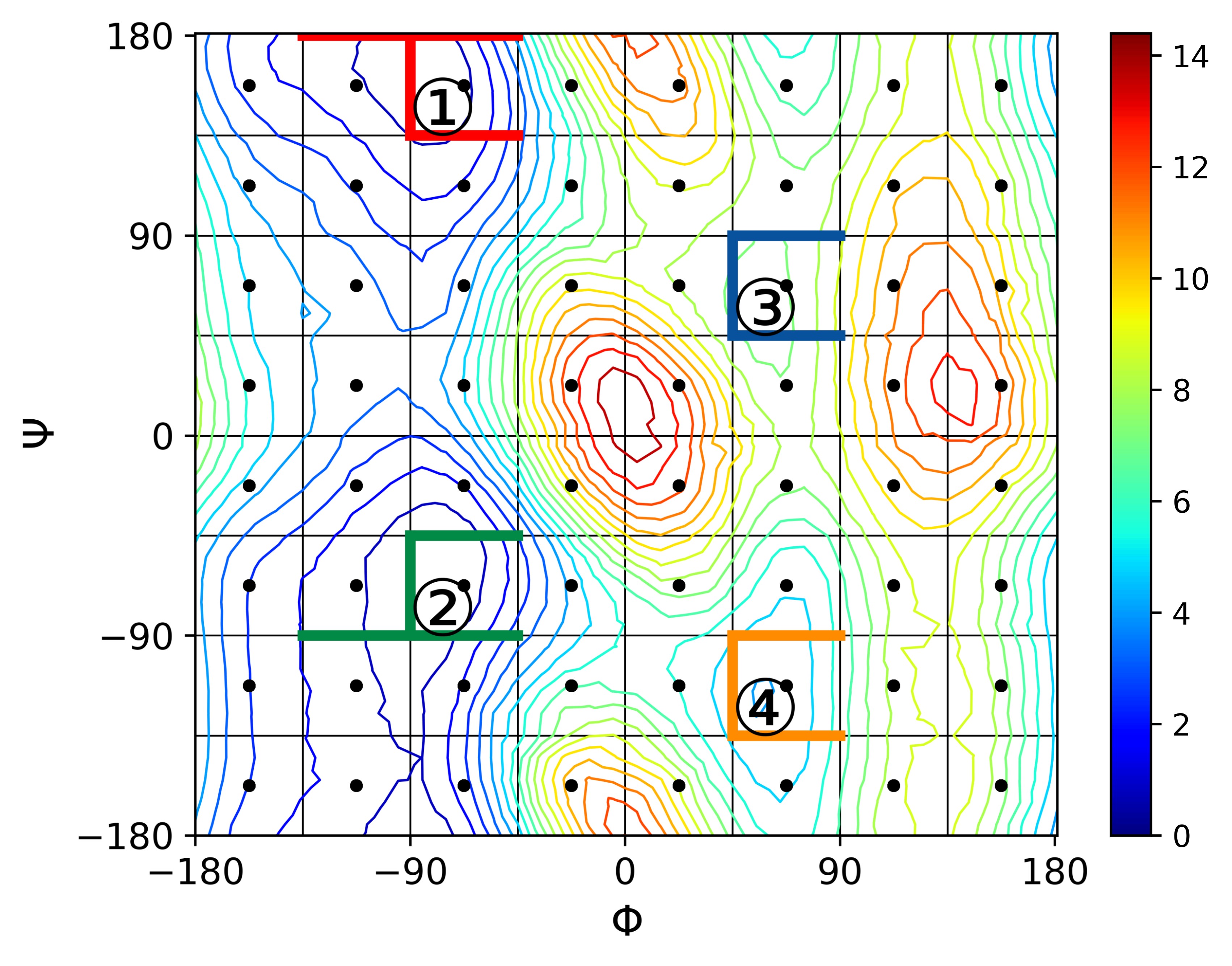}\\
(a)}
\makecell{
\includegraphics[height=5cm]{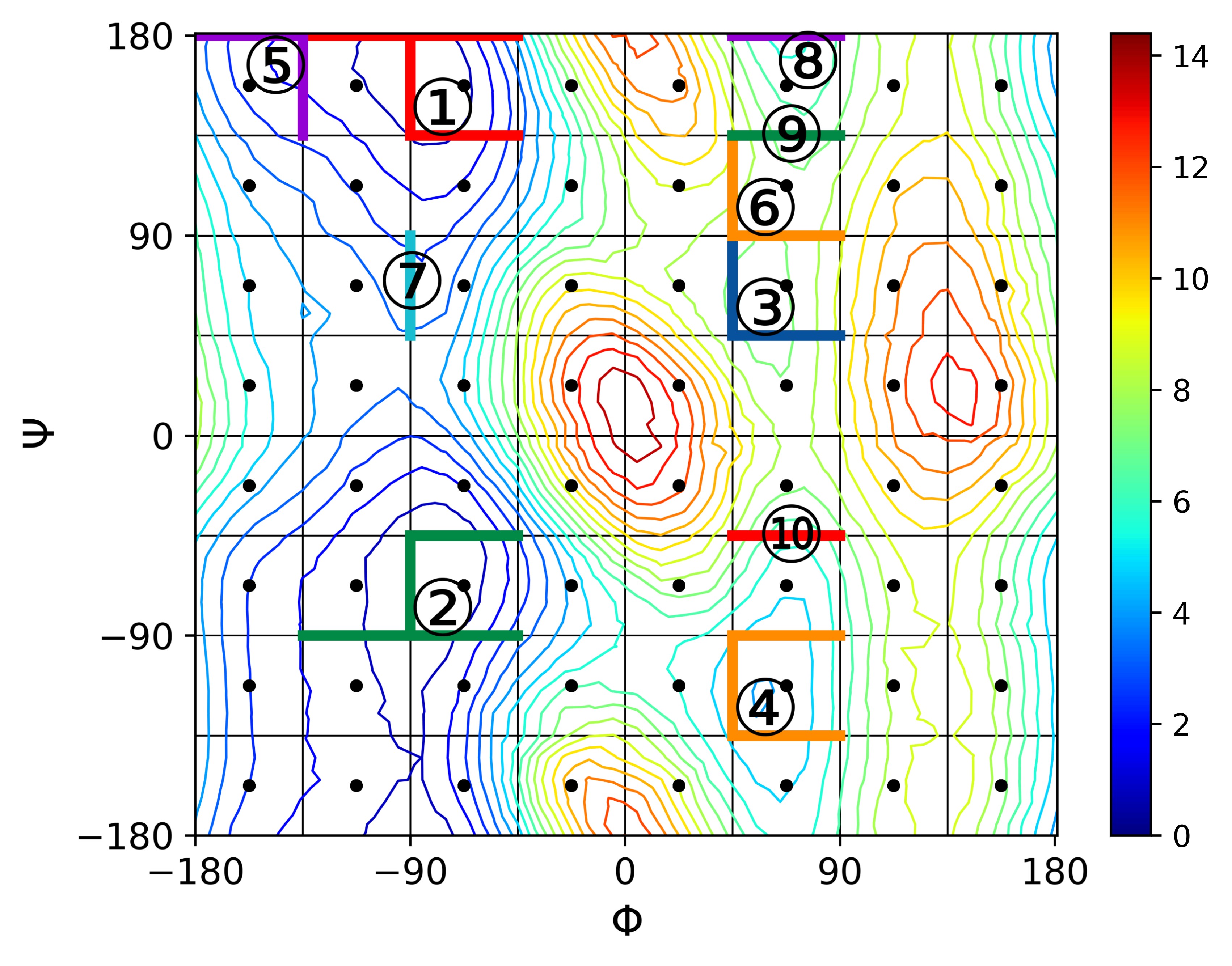} \\
(d)}\\
\makecell{
\includegraphics[height=4.5cm]{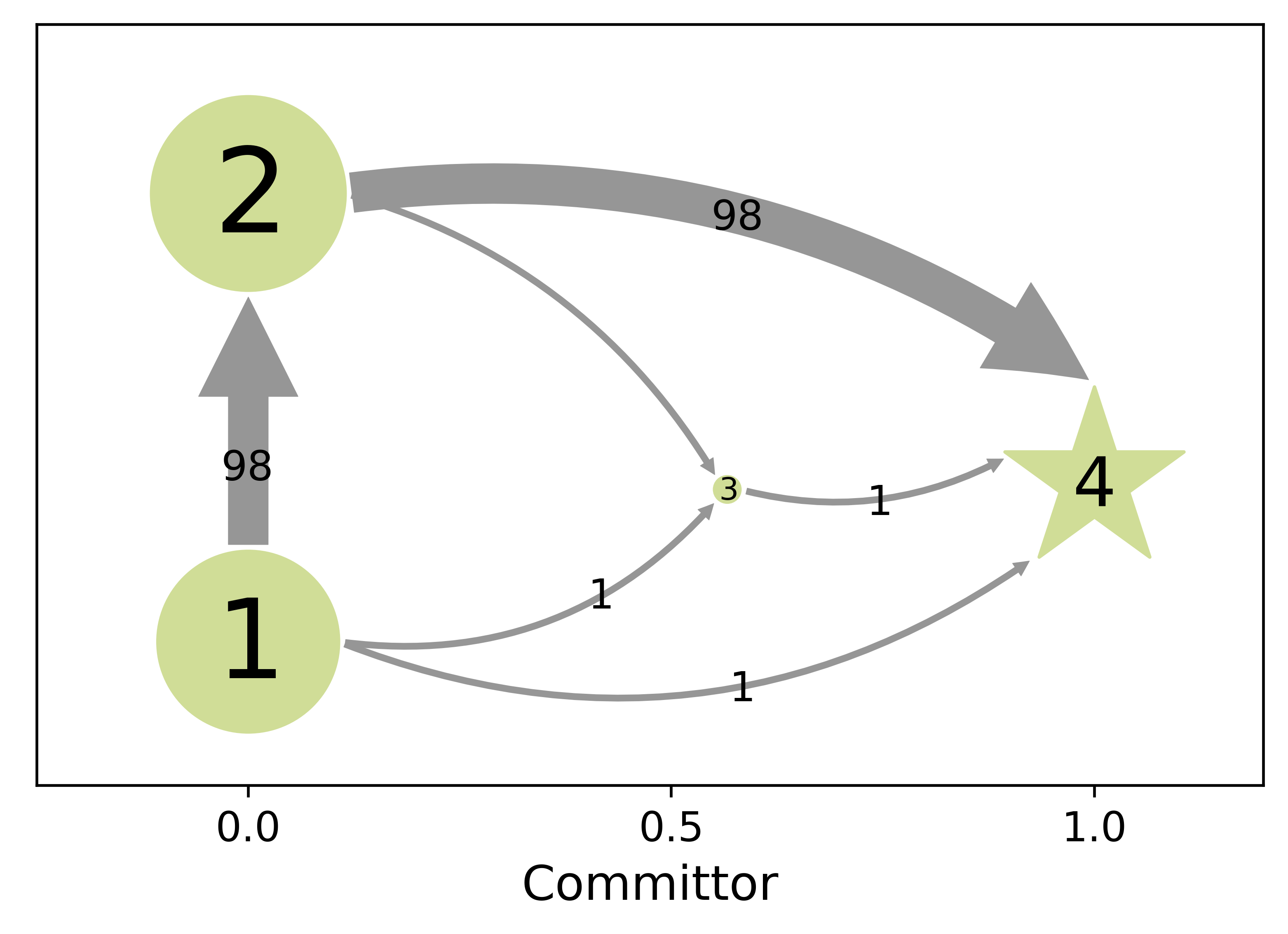}\\
(b)}
\makecell{
\includegraphics[height=4.5cm]{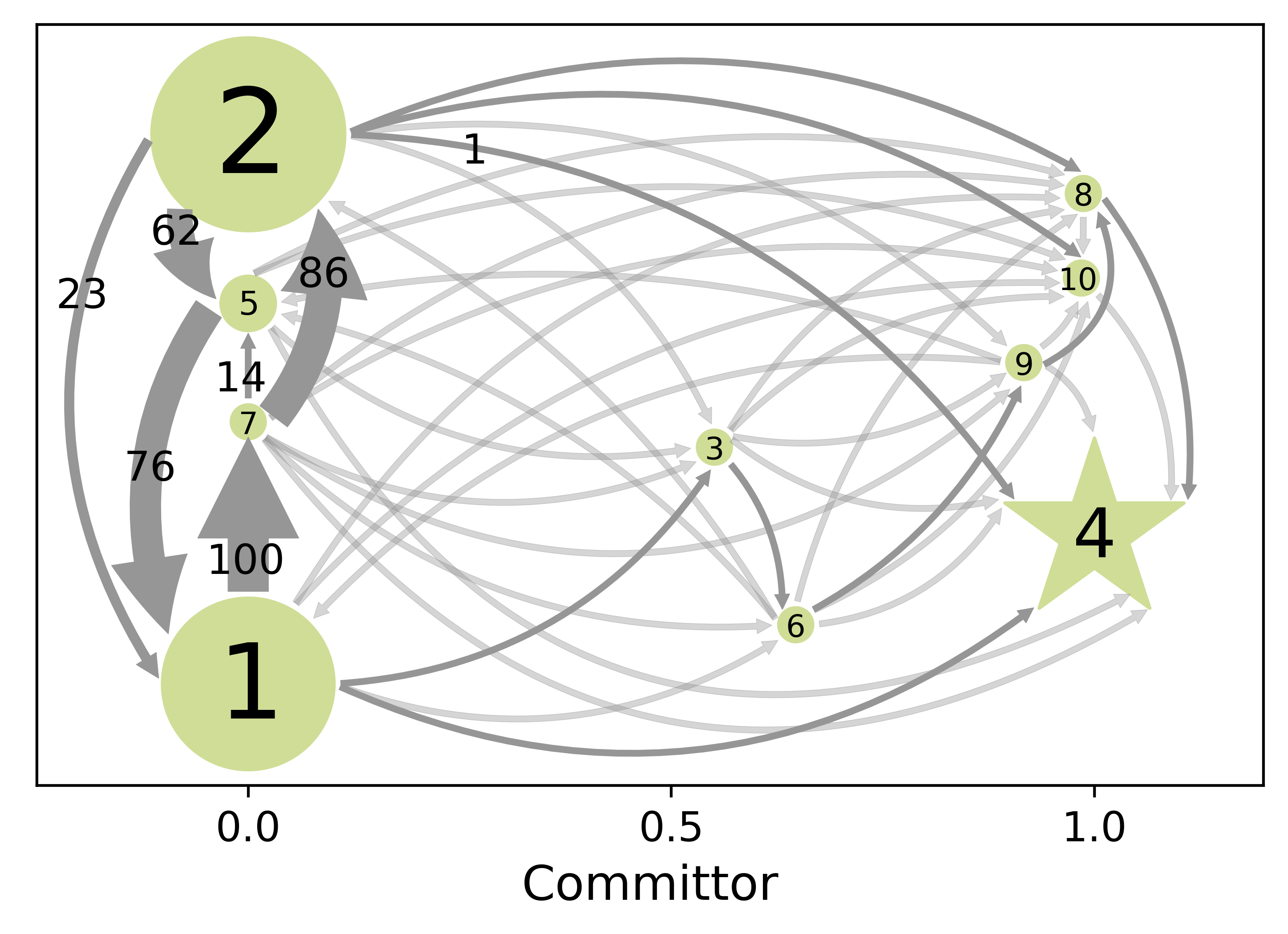} \\
(e)} \\
\makecell{
\includegraphics[height=4cm]{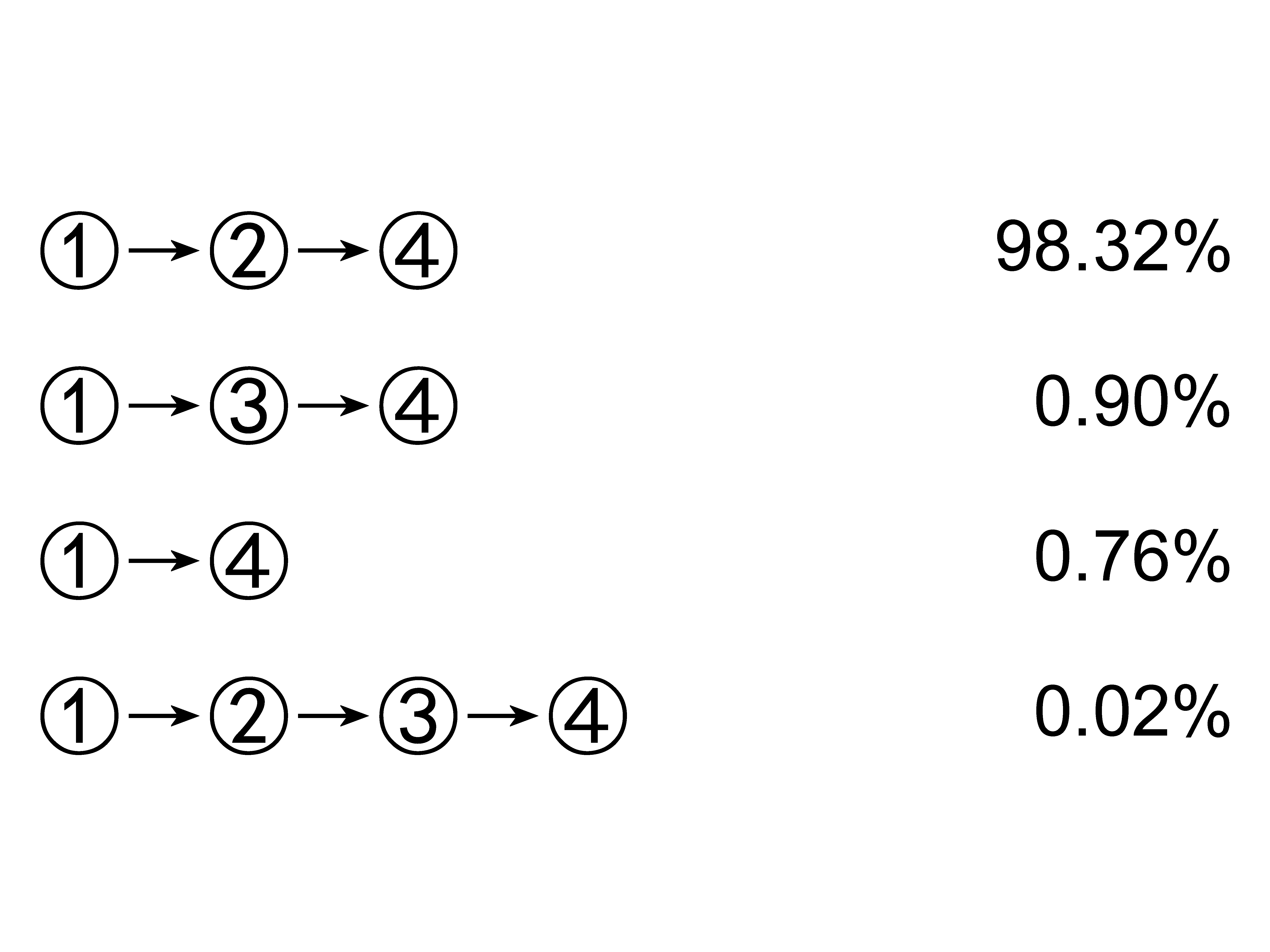}\\
(c)}
\makecell{
\includegraphics[height=4cm]{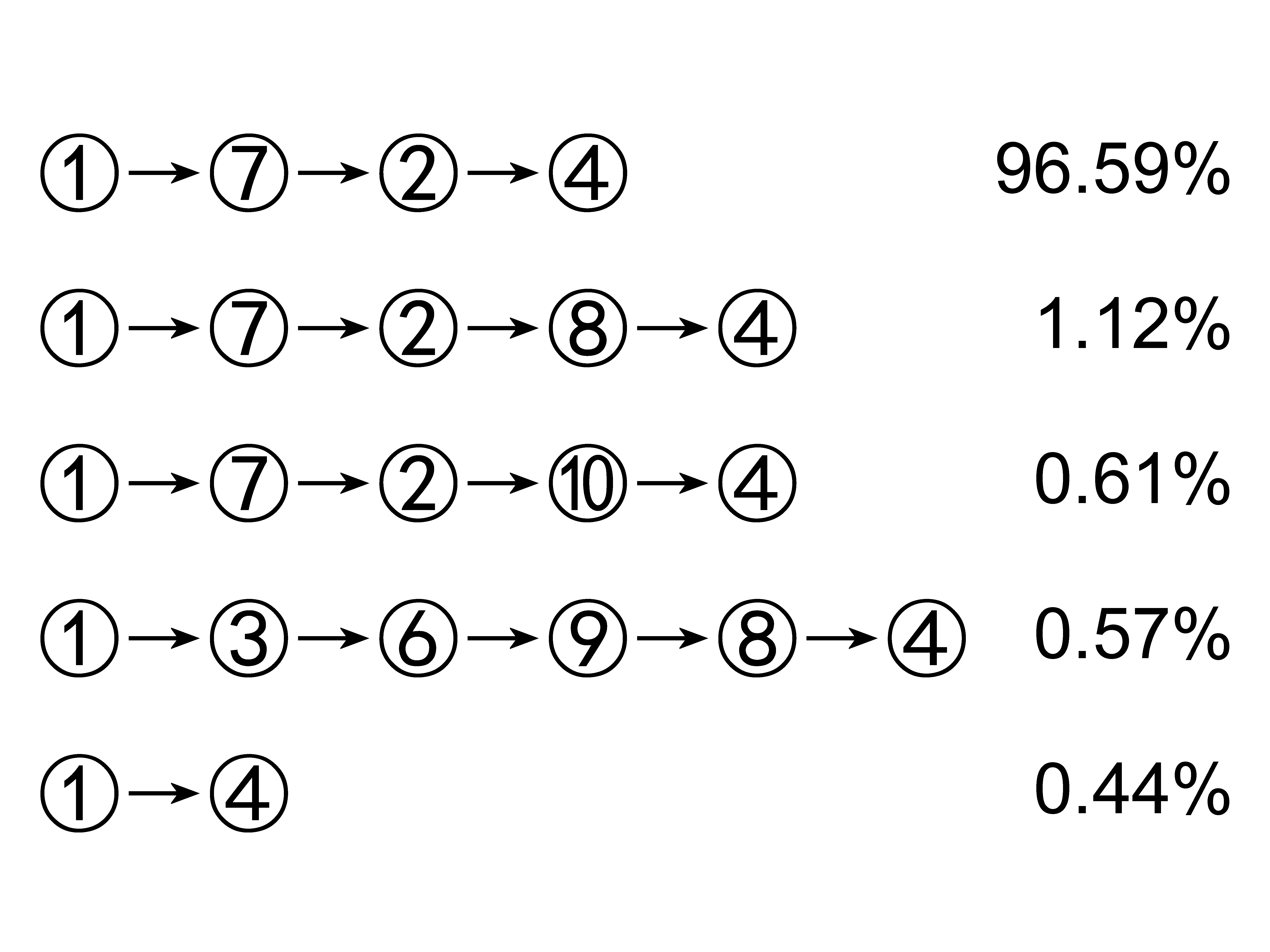} \\
(f)}\\
\end{tabular}
\caption{Clustering and reduction results of the two-angle calculation with a solvated alanine dipeptide. The left panel ((a)-(c)) and right panel ((d)-(e)) show the four-cluster and ten-cluster results, respectively. Top: Milestones are grouped into clusters (shown in different colors). The core set of each cluster is indicated by bold lines. The contour plot represents the free energy landscape (in kcal/mol) obtained from the combined umbrella sampling calculations and weighted histogram analysis method. Middle: The reduced network with normalized effective current (rounded off to integers). The total outflow current from the reactant state (\ding{172}) is set to 100. Effective currents smaller than 1 are not labeled. The size of each node (circle) is proportional to the stationary probability of the corresponding core set. The product state (\ding{175}) is marked as a star. Bottom: The first few dominant transition pathways and their contribution ratios are listed.}\label{fig_ala2_red}
\end{figure}

The spectral structure of the composite matrix $\mathbf{A}$ is shown in Fig. \ref{fig_ala2_spec} (b), indicating four main metastable states.
The identified core sets (\ding{172}-\ding{175} in Fig. \ref{fig_ala2_red} (a)) align with the four main free energy minima on the free energy landscape.
The core sets of the first cluster and fourth cluster are designated as the reactant state (\ding{172}, $\mathrm{C_{5}}$ configuration) and the product state (\ding{175}, $\mathrm{C7_{ax}}$ configuration), respectively.
The resulting reduced network contains only four nodes (Fig. \ref{fig_ala2_red} (b)), providing a simplified representation of the transition process.
The first few dominant transition pathways are listed in Fig. \ref{fig_ala2_red} (c).
The transition path via the $\alpha_R$ configuration, \ding{172}$\rightarrow$\ding{173}$\rightarrow$\ding{175}, dominates and contributes over 98\%.
The next dominant transition path is via the $\alpha_L$ configuration, \ding{172}$\rightarrow$\ding{174}$\rightarrow$\ding{175}.
In this example of the alanine dipeptide with periodic boundary conditions in $\phi$ and $\psi$, even transition paths between two adjacent core sets are not unique.
For example, in the four-cluster case, it is unclear which specific path it takes to transit between core sets \ding{174} and \ding{175} in the second dominant pathway.
This would require adding more intermediate states.

The next spectral gap occurs between the tenth and eleventh eigenvalues of $\mathbf{A}$.
The resulting reduced network has ten nodes (Fig. \ref{fig_ala2_red} (d)), which provides a more detailed description of the transition process among the four main metastable states.
The ten nodes can be roughly classified into two categories based on their committor values (Fig. \ref{fig_ala2_red} (e)).
The four nodes (\ding{172}, \ding{173}, \ding{176}, and \ding{178}) in the range $\phi\in[-180\degree, 0\degree]$ are more likely to first reach the reactant state (\ding{172}) before the product state (\ding{175}).
These four nodes have a strong interconnection among themselves.
The other six nodes in the range $\phi\in[0\degree, 180\degree]$ are more likely to first reach the product state.
The effective current between these two categories of nodes is relatively small, which is consistent with the free energy landscape.
Now, the transition paths among four main metastable states become clearer.
For example, the dominant transition path between \ding{174} and \ding{175} can now be identified as the one traversing through $\psi=180\degree$ (Fig. \ref{fig_ala2_red} (f)).

\subsubsection{Four-angle Calculation}
It has been demonstrated that, for a more comprehensive characterization of the alanine dipeptide isomerization, more degrees of freedom than $\phi$ and $\psi$ are needed\cite{RCala2}.
Hence, we incorporate two additional dihedral angles $\theta$ and $\zeta$ into our analysis\cite{SSFE}.
The Milestoning network generated from four-angle partitioning represents a more challenging test, because (i) there are much more milestones, and (ii) the free energy landscape in four angles is impossible to visualize.

\begin{figure}[h]
\centering
\begin{tabular}{cccccc}
\makecell{
\includegraphics[height=5.5cm]{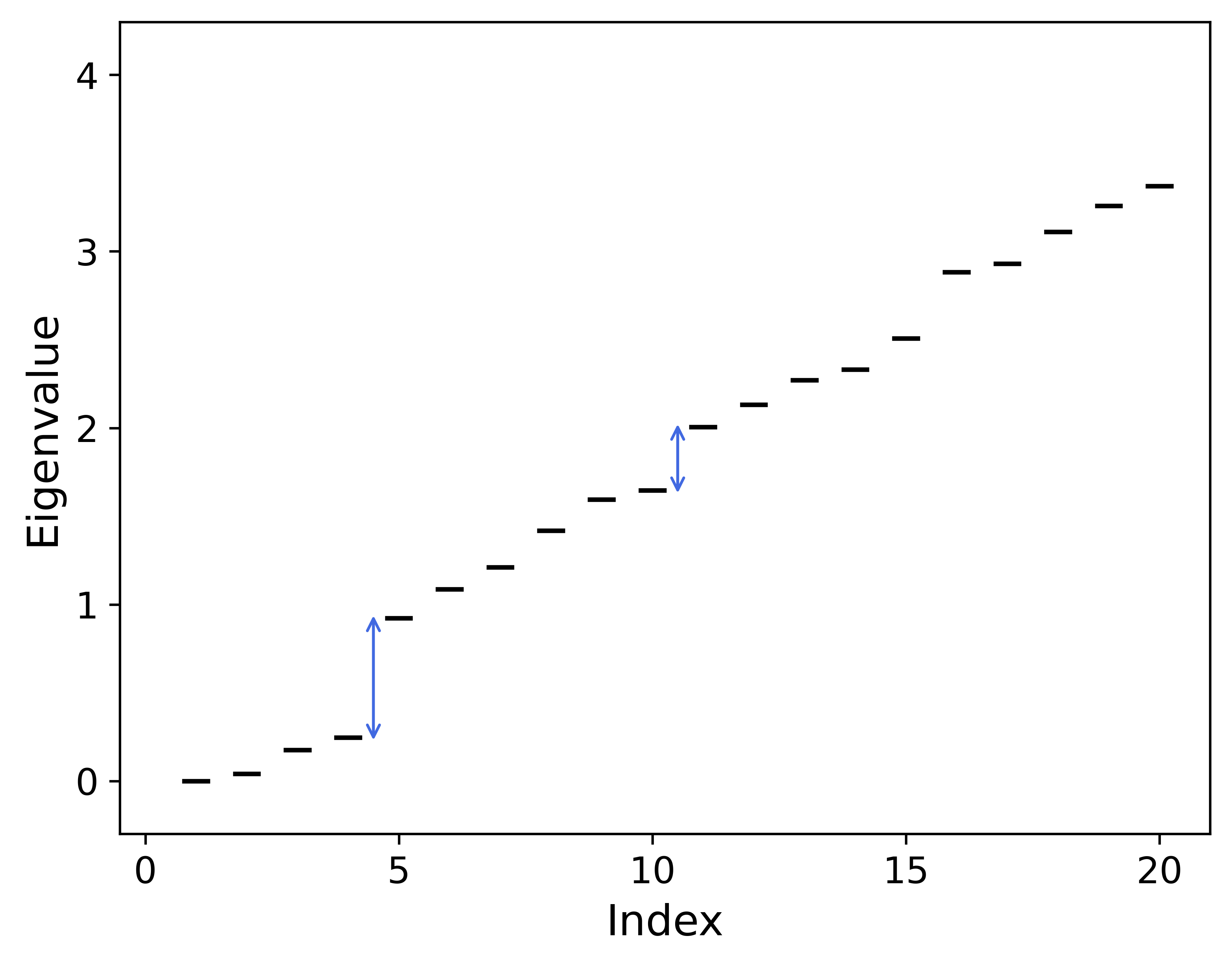} \\
(a)} \\
\makecell{
\includegraphics[height=5cm]{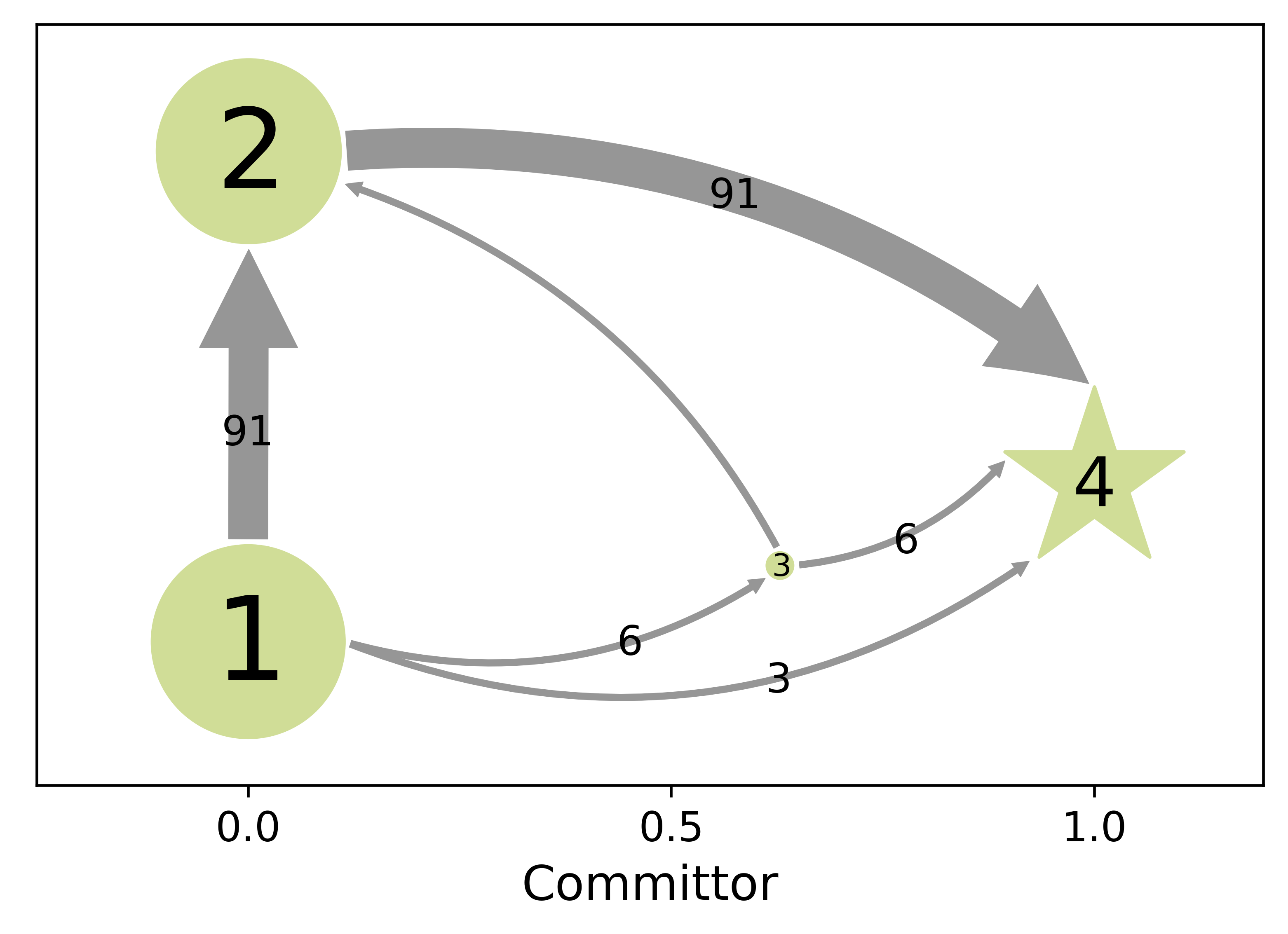} \\
(b)} \\
\makecell{
\includegraphics[height=4.5cm]{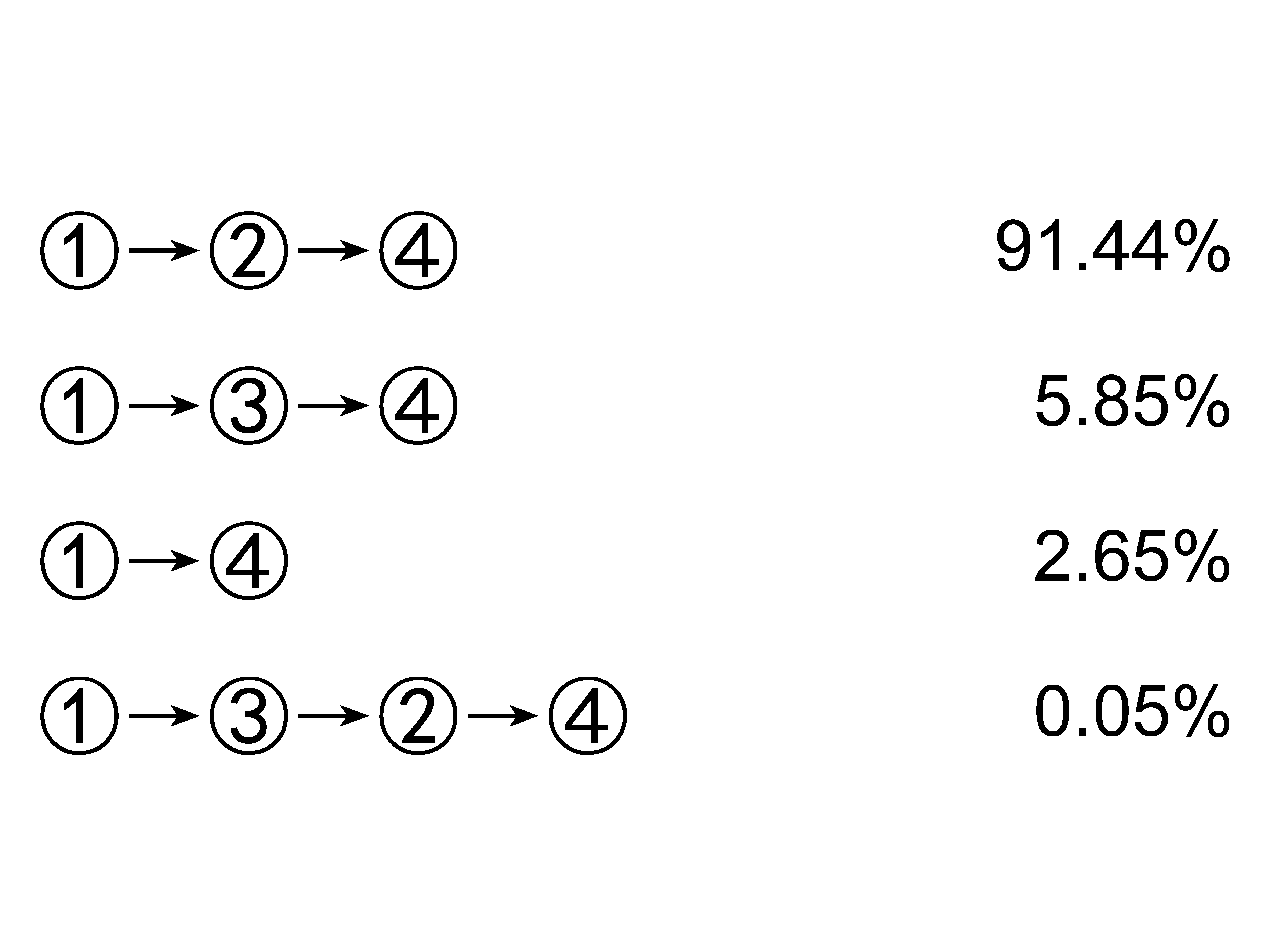} \\
(c)} \\
\end{tabular}
\caption{The four-angle calculation with a solved alanine dipeptide. (a) The eigenvalues of the composite matrix $\mathbf{A}=\mathbf{Q}\tilde{\mathbf{Q}}$ in ascending order. The first two spectral gaps are indicated by arrows. (b) The reduced network with normalized effective current (rounded off to integers). The total outflow current from the reactant state (\ding{172}) is set to 100. Effective currents smaller than 1 are not labeled. The size of each node (circle) is proportional to the stationary probability of the corresponding core set. The product state (\ding{175}) is marked as a star. (c) The first few dominant transition pathways and their contribution ratios are listed.}\label{fig_ala4_red}
\end{figure}

The spectral structure of the composite matrix $\mathbf{A}$ (Fig. \ref{fig_ala4_red} (a)) exhibits a similar pattern to that of two-angle calculations, indicating the presence of four main metastable states.
Upon projecting the core sets of these four clusters back onto $(\phi, \psi)$ space, we find them to be located in the same regions as those in the two-angle calculations.
Once again, the core sets \ding{172} and \ding{175} are designated as the reactant state and the product state, respectively.
Although the effective currents among core sets are slightly different from those in the two-angle calculations (Fig. \ref{fig_ala4_red} (b)), the order of the first few dominant transition pathways remains the same (Fig. \ref{fig_ala4_red} (c)).

As a more stringent test, the $512\times512$ transition probability and transition time matrices generated in the four-angle calculation are reduced into matrices of size $128\times128$ with the same set of nodes used in the two-angle calculations.
This involves recombining the fine division of four cells in $(\theta, \zeta)$ into one milestone in $(\phi, \psi)$.
Remarkably, the reduced transition probabilities and average residence time agree well with those directly obtained in the two-angle calculation.
The largest deviation in transition probabilities and average residence time is around 0.1 and 0.08 ps, respectively (data not shown). 

The analysis above shows that the confinement of $\theta$ and $\zeta$ in the range $[-60\degree, 60\degree]$ does not significantly influence on the isomerization process of \ding{172}$\rightarrow$\ding{175} for the solved alanine dispeptide.

\section{Conclusion}\label{Conclusion}
In summary, we have introduced an accurate and efficient reduction analysis algorithm, dubbed ReM, for the Milestoning network, which involves three sequential steps: clustering, reduction and transition path analysis.
The reduction process has resolved three challenges: (i) the non-Markovian nature of the Milestoning formulation; (ii) potential violation of time-reversibility in the network generated from practical simulations; (iii) preservation of kinetic properties during reduction.
The first of these is addressed by introducing an auxiliary CTMC, which shares the same stationary probability, local residence time and MFPT between any two milestones as the Milestoning formulation.
Leveraging this auxiliary CTMC, we introduce a composite matrix to address the complex-value issue of the non-reversible CTMC (or equivalently, the non-reversible Milestoning network, which is the second challenge). 
Due to the nice properties of the introduced composite matrix, its eigenvectors contain characteristic information of metastable states and are therefore used for clustering.
The third challenge pertains to the reduction step.
Through proper network transformation and recalculation of transition probabilities and transition time, kinetic properties such as local residence time, exit time and the MFPT between any two states (designated as the reactant and product state, respectively) are preserved.
Compared to the direct analysis of reaction coordinates and pathways in the original high-dimensional Milestoning network, our new method provides a simplified yet comprehensive interpretation of rare transition processes in complex systems.
Furthermore, by tuning the truncation place in the spectra of the composite matrix, the complexity of the reduced network can be tailored to meet various needs.

\clearpage
\newpage

\begin{algorithm}[H]
\caption{The reduction of Milestoning (ReM) algorithm.}
\label{ReM}
\begin{algorithmic}[1]
\Inputs {$\mathbf{K}\gets$ Eq. \eqref{K simu} \Comment{The transition probability matrix in the original network}\\
$\mathbf{T}\gets$ Eq. \eqref{T simu} \Comment{The transition time matrix in the original network}}
\State // Begin clustering //
\State $\mathbf{t}\gets \mathbf{T}\mathbf{1}$\Comment{Calculate the residence time on each milestone}
\State $Q_{ba}=K_{ba}/t_b$ (for $a\neq b$), $Q_{bb}=-\sum_{a\neq b}Q_{ba}$ \Comment Construct an auxiliary transition rate matrix
\State $\boldsymbol{\Pi}\gets\mathbf{\pi}^T\mathbf{Q}=\mathbf{0}^T$ \Comment{Solve for stationary probabilities}
\State $\tilde{\mathbf{Q}}\gets \boldsymbol{\Pi}^{-1}\mathbf{Q}^T\boldsymbol{\Pi}$ \Comment{Construct the time-reversed transition rate matrix $\tilde{\mathbf{Q}}$}
\State $\mathbf{A}\gets\mathbf{Q}\tilde{\mathbf{Q}}$ \Comment{Construct the composite matrix $\mathbf{A}$}
\State $(\mathbf{u}_1,\mathbf{u}_2,\cdots,\mathbf{u}_k)\gets \mathbf{A}\mathbf{u}_i=\lambda_i\mathbf{u}_i$\Comment{The first $k$ low-lying eigenvectors truncated at spectral gaps}
\State $k$ clusters $\gets$ $k$-means clustering in $(\mathbf{u}_2,\cdots,\mathbf{u}_k)$
\State // End clustering //
\State // Begin reduction //
\State $\{C_1,\cdots,C_k\}\gets$ Eq. \eqref{core sets criterion}\Comment{Identify core sets of the $k$ clusters}
\State $C_i\gets$ Reactant, $C_f\gets$ Product\Comment{Designate the reactant state and the product state}
\State $\mathbf{K}^{(I)}\gets$ Eq. \eqref{K inter}\Comment{Calculate the transition probability matrix in the intermediate network}
\State $\mathbf{T}^{(I)}, \mathbf{t}^{(I)}\gets$ Eqs. \eqref{T inter} and \eqref{t inter}\Comment{Calculate the transition time in the intermediate network}
\State $\mathbf{K}^{(R)}\gets$ Eq. \eqref{K reduced} \Comment{Calculate the transition probability matrix in the fully reduced network}
\State $\mathbf{T}^{(R)}, \mathbf{t}^{(R)}\gets$ Eqs. \eqref{T reduced} and \eqref{t reduced} \Comment{Calculate the transition time in the fully reduced network}
\State // End reduction //
\State // Begin transition path analysis //
\State $\{E_{C_aC_b}\}\gets$ Eq. \eqref{edge weight}\Comment{Define directed edge weights in the fully reduced network}
\State Find the GMWP using the recursive Dijkstra's algorithm
\While {The network is connected from $C_i$ to $C_f$}
\State Adjust the edge weights according to Eq. \eqref{edge weight adjust}
\State Find the next GMWP in the updated network
\EndWhile
\State $R(\omega^{(i)}_{GMWP})\gets$ Eq. \eqref{path ratio}\Comment{Calculate the contribution ratio of each GMWP}
\State // End transition path analysis //
\end{algorithmic}
\end{algorithm}

\begin{acknowledgments}
The work was partially supported by Qilu Young Scholars Program of Shandong University and Natural Science Foundation of Shandong Province (No. ZR2022QA012).
\end{acknowledgments}

\section*{Data Availability Statement}
The data that support the findings of this study are available within the article and its supplementary material.

\section*{Conflicts of interest}
There are no conflicts to declare.


\clearpage
\newpage

\appendix
\section{Properties of the composite matrix $\mathbf{A}=\mathbf{Q}\tilde{\mathbf{Q}}$}\label{app1}
\begin{prop} 
The composite matrix $\mathbf{A}$ satisfies the detailed balance condition.
\end{prop}

\begin{proof}
For each $a, b\in\mathcal{M}$, we have
\begin{align}
\pi_aA_{ab} &= \sum_{c\in\mathcal{M}}\pi_aQ_{ac}\tilde{Q}_{cb}\nonumber \\
&=\sum_{c\in\mathcal{M}}\pi_aQ_{ac}\frac{\pi_b}{\pi_c}Q_{bc}\nonumber \\
&= \sum_{c\in\mathcal{M}}\tilde{Q}_{ca}\pi_bQ_{bc}\nonumber \\
&=A_{ba}\pi_b.
\end{align}
\end{proof}

\begin{prop}
The eigenvalues of the composite matrix $\mathbf{A}$ are real and non-negative.
\end{prop}

\begin{proof}
Define a similarity transform of the composite matrix $\mathbf{A}$ by 
\begin{align}
\mathbf{B} &= \boldsymbol{\Pi}^{1/2}\mathbf{A}\boldsymbol{\Pi}^{-1/2}\nonumber\\
&= \boldsymbol{\Pi}^{1/2}\mathbf{Q}\tilde{\mathbf{Q}}\boldsymbol{\Pi}^{-1/2}\nonumber\\
&=\boldsymbol{\Pi}^{1/2}\mathbf{Q}\boldsymbol{\Pi}^{-1}\mathbf{Q}^T\boldsymbol{\Pi}^{1/2}\nonumber\\
&=(\boldsymbol{\Pi}^{1/2}\mathbf{Q}\boldsymbol{\Pi}^{-1/2})(\boldsymbol{\Pi}^{-1/2}\mathbf{Q}^T\boldsymbol{\Pi}^{1/2})\nonumber\\
&\equiv\mathbf{R}\mathbf{R}^T,
\end{align}
where going from the second line to the third we have used the definition of $\tilde{\mathbf{Q}}$ in Eq. \eqref{q tilde}.
The matrix $\mathbf{R}$ can be further decomposed via the singular value decomposition, $\mathbf{R}=\mathbf{U}\boldsymbol{\Sigma}\mathbf{V}^T$. Finally, we obtain
\begin{equation}
\mathbf{B}=\mathbf{U}\boldsymbol{\Sigma}^2\mathbf{U}^T,
\end{equation}
where $\boldsymbol{\Sigma}$ is a diagonal matrix containing the singular values of $\mathbf{R}$,
\begin{equation}
\boldsymbol{\Sigma} = \begin{pmatrix}
\sigma_1 & 0 & \cdots & \cdots & 0\\
0 & \sigma_2 & 0 & \cdots & 0\\
0 & 0 & \sigma_3 & \cdots & 0\\
\vdots & \vdots & \vdots & \ddots &\vdots\\
0 & \cdots & \cdots & 0 & \sigma_N\\
\end{pmatrix}.
\end{equation}
Since matrices $\mathbf{B}$ and $\mathbf{A}$ are connected by a similarity transform, they share the same eigenvalues. Consequently, the eigenvalues of the composite matrix $\mathbf{A}$ are related to the singular values of matrix $\mathbf{R}$ via $\lambda_i=\sigma_i^2\geq 0$.
\end{proof}

It is noteworthy that for a time-reversible CTMC, matrix $\mathbf{R}$ is symmetric, i.e., $\mathbf{R}=\mathbf{R}^T$. 
It follows that the eigenvalue $\mu_i$ and singular value $\sigma_i$ of matrix $\mathbf{R}$ are related by $\mu_i^2=\sigma_i^2$. 
Since $\mathbf{R}$ and $\mathbf{Q}$ are connected by a similarity transformation, they have the same eigenvalues.
As a result, the eigenvalue of the composite matrix $\mathbf{A}$ is related to that of $\mathbf{Q}$ via $\lambda_i=\mu_i^2$.

\section{Network transformation operations preserving MFPT}\label{app2}
The reduction from the original network to the intermediate network involves two basic operations: (i) removing milestones not belonging to core sets; (ii) setting milestones within the same core set invisible to each other.
Following these two operations, the transition probabilities and transition time should be adjusted accordingly as well.

In the Milestoning formulation, the MFPT from the reactant (milestone $r$) to the product state (milestone $p$) is calculated by
\begin{equation}
\tau_{r\rightarrow p}=\mathbf{e}_r^T(\mathbf{I}-\mathbf{K}'')^{-1}\mathbf{t}'',
\label{MFPT eq}
\end{equation}
Here, $\mathbf{K}''$ and $\mathbf{t}''$ are different from the original $\mathbf{K}$ and $\mathbf{t}$ only at the product state, where absorbing boundary conditions are now imposed, $K''_{pa}=0, \forall a\in\mathcal{M}$, and $t''_p=0$.
Equation \eqref{MFPT eq} indicates that if the transition probabilities ($\mathbf{K}$) and average residence time ($\mathbf{t}$) are "equivalent" before and after a network transformation, the MFPTs will be the same.
In the following, we will show that the proper adjustment of $\mathbf{K}$ and $\mathbf{t}$ is given by Eqs. \eqref{K inter} and \eqref{t inter}, respectively.

\begin{figure}[h]
\centering
\begin{tabular}{cccccc}
\includegraphics[height=8cm]{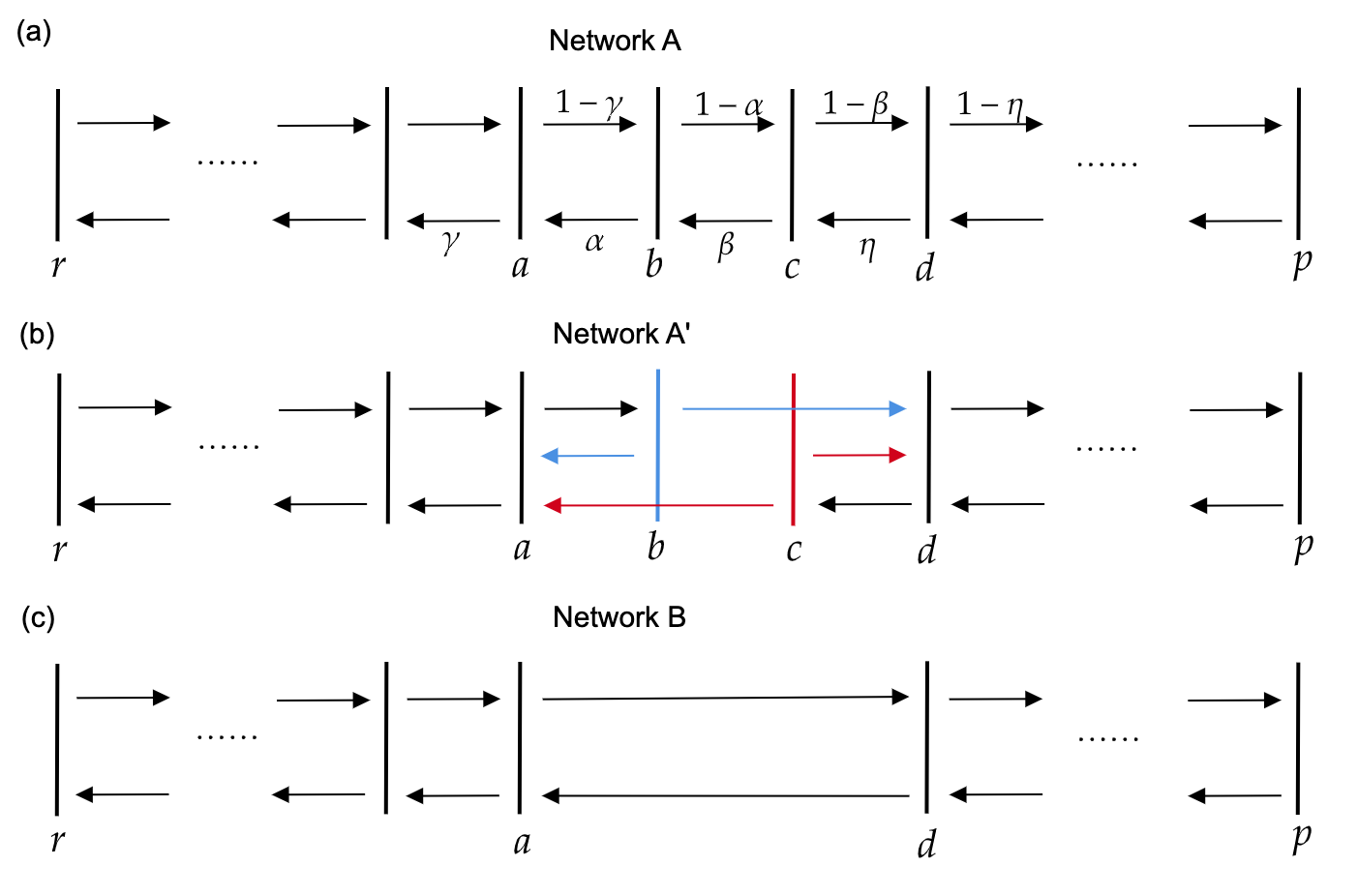} \\ 
\end{tabular}
\caption{(a) A one-dimensional network $A$. Each milestone can only communicate with its two neighbors. Transition probabilities are labeled next to arrows. (b) Network $A'$: two milestones $b$ and $c$ in Network $A$ are set invisible to each other. The transition probabilities and transition time associated with milestones $b$ and $c$ are adjusted. (c) Two milestones $b$ and $c$ are removed from Network $A$ and $A'$. Transition probabilities and transition time associated with milestones $a$ and $d$ are adjusted.}\label{fig_appen}
\end{figure}

The fact that the operation (i) (cf. Fig. \ref{fig_appen} (a)$\rightarrow$(c)) along with the corresponding adjustment (Eqs. \eqref{K inter} and \eqref{t inter}) preserves MFPT has been demonstrated by Wales and co-workers\cite{GT06,GT09}.
The proof was conducted in the case of a CTMC by explicitly summing up all possible transition paths.
According to the equivalence between a CTMC induced from $\mathbf{K}$ and $\mathbf{t}$ and the Milestoning network in MFPT calculation (cf. Eq. \eqref{MFPT eq}), the operation (i) performed in the Milestoning network also preserves MFPT.

We therefore focus on the operation (ii).
For simplicity, a one-dimensional network is considered.
But the conclusion can be generalized to networks of higher dimensions.

Consider four consecutive intermediate milestones in Network $A$ (cf. Fig. \ref{fig_appen} (a) $a$-$d$).
The associated transition probabilities and average residence time are given by
\begin{equation}
\mathbf{K}^{(A)} = \begin{pmatrix}
\cdots & \cdots & \cdots   & \cdots & \cdots & \cdots\\
\cdots & 0      & 1-\gamma & 0      & 0      & \cdots\\
\cdots & \alpha & 0 & 1-\alpha & 0 &\cdots\\
\cdots & 0 & \beta & 0 & 1-\beta &\cdots\\
\cdots & 0 & 0 & \eta & 0 & \cdots \\
\cdots & \cdots &\cdots & \cdots & \cdots &\cdots
\end{pmatrix}\begin{matrix}
\ \\
\rightarrow a\\
\rightarrow b\\
\rightarrow c\\
\rightarrow d\\
\ \\
\end{matrix},
\label{appen K A}
\end{equation}
and 
\begin{equation}
\mathbf{t}^{(A)}=\begin{pmatrix}
\cdots\\
t_a\\
t_b\\
t_c\\
t_d\\
\cdots
\end{pmatrix}\begin{matrix}
\ \\
\rightarrow a\\
\rightarrow b\\
\rightarrow c\\
\rightarrow d\\
\ \\
\end{matrix},
\label{appen t A}
\end{equation}
respectively. 

Now two milestones ($b$ and $c$) are set invisible to each other, resulting in a transformed network $A'$ (Fig. \ref{fig_appen} (b)).
Applying Eqs. \eqref{K inter} and \eqref{t inter} for updating transitions initiated from $b$ and $c$, the resulting new transition probabilities and average residence time are given by
\begin{equation}
\mathbf{K}^{(A')} = \begin{pmatrix}
\cdots & \cdots & \cdots   & \cdots & \cdots & \cdots\\
\cdots & 0   & 1-\gamma & 0      & 0      & \cdots\\
\cdots & \frac{\alpha}{1+(\alpha-1)\beta} & 0 & 0 & \frac{(1-\alpha)(1-\beta)}{1+(\alpha-1)\beta} &\cdots\\
\cdots & \frac{\alpha\beta}{1+(\alpha-1)\beta} & 0 & 0 & \frac{1-\beta}{1+(\alpha-1)\beta} &\cdots\\
\cdots & 0 & 0 & \eta & 0 & \cdots \\
\cdots & \cdots &\cdots & \cdots & \cdots &\cdots
\end{pmatrix}\begin{matrix}
\ \\
\rightarrow a\\
\rightarrow b\\
\rightarrow c\\
\rightarrow d\\
\ \\
\end{matrix},
\end{equation}
and
\begin{equation}
\mathbf{t}^{(A')}=\begin{pmatrix}
\cdots\\
t_a\\
\frac{t_b+(1-\alpha)t_c}{1+(\alpha-1)\beta}\\
\frac{\beta t_b+t_c}{1+(\alpha-1)\beta}\\
t_d\\
\cdots
\end{pmatrix}\begin{matrix}
\ \\
\rightarrow a\\
\rightarrow b\\
\rightarrow c\\
\rightarrow d\\
\ \\
\end{matrix},
\end{equation}
respectively. 
The "$\cdots$" indicates corresponding elements from Eqs. \eqref{appen K A} and \eqref{appen t A}.

Finally, milestones $b$ and $c$ are removed from Network $A'$ (operation (i)), resulting in Network $B$ (Fig. \ref{fig_appen} (c)).
Again, Eqs. \eqref{K inter} and \eqref{t inter} are applied to update the transition probabilities and average residence time associated with milestones $a$ and $d$.
After some algebra, it can be readily verified that the two transformation pathways ($A\stackrel{\mathrm{(i)}}\longrightarrow B$ and $A\stackrel{\mathrm{(ii)}}\longrightarrow A'\stackrel{\mathrm{(i)}}\longrightarrow B$) lead to the same set of $\mathbf{K}^{(B)}$ and $\mathbf{t}^{(B)}$ in Network $B$.

Since the operation (i) preserves the MFPT, we have $\tau_{r\rightarrow p}^{(A)}=\tau_{r\rightarrow p}^{(B)}$ and $\tau_{r\rightarrow p}^{(A')}=\tau_{r\rightarrow p}^{(B)}$.
Therefore, we arrive at the final conclusion $\tau_{r\rightarrow p}^{(A)}=\tau_{r\rightarrow p}^{(A')}$.

\clearpage
\newpage

\bibliographystyle{achemso1}
\bibliography{MReduction}

\clearpage
\newpage

TOC
\begin{figure}[h]
\centering
\begin{tabular}{cc}
\includegraphics[height=7cm]{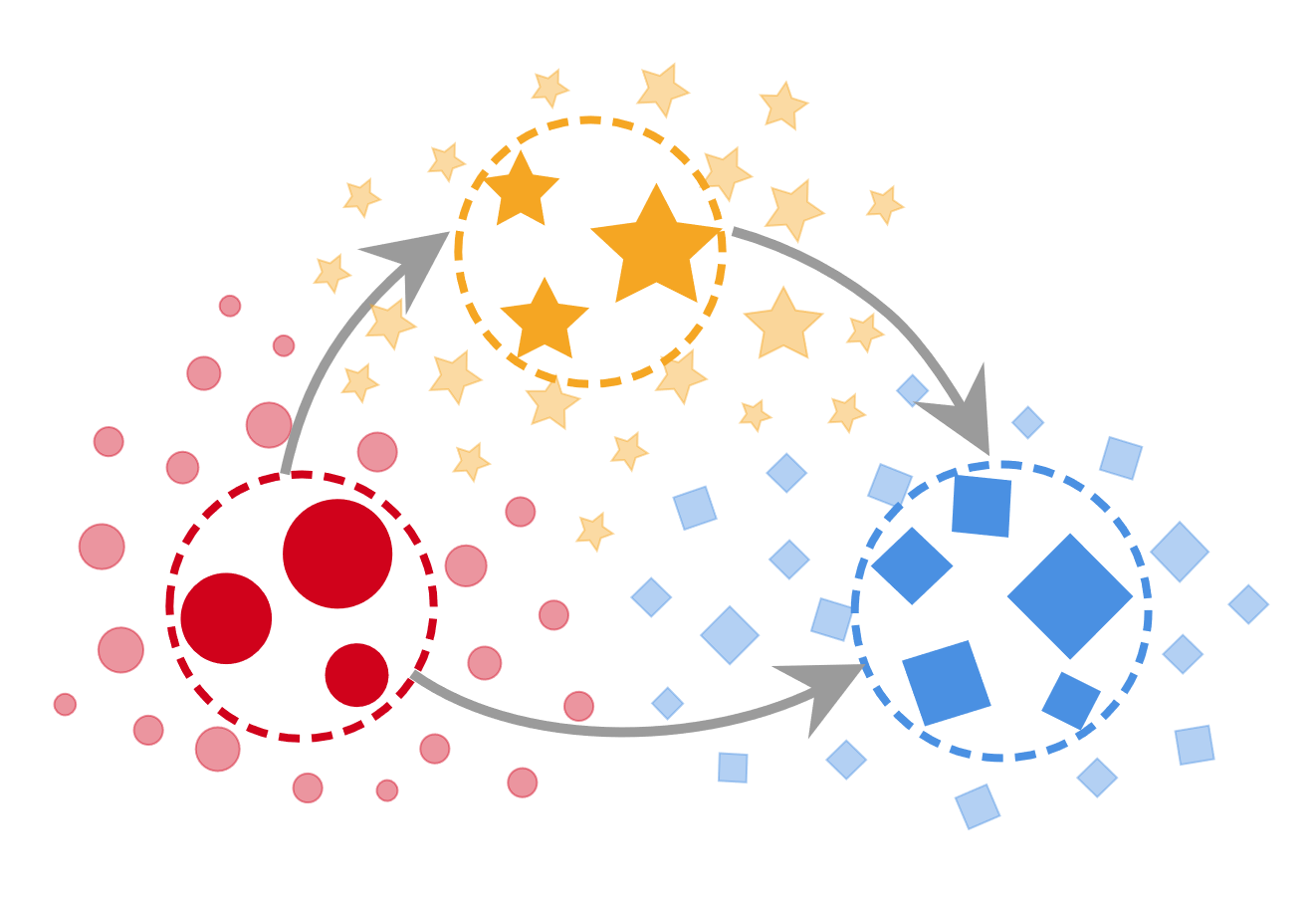}\\
\end{tabular}
\end{figure}

\end{document}